\documentclass{article}

\usepackage{fullpage}
\usepackage{color}
\usepackage{lambda,ifthen,calc}
\usepackage{tabularx}
\usepackage{fancyvrb}
\usepackage[final]{showexpl}

\usepackage[inactive]{pst-pdf}
\usepackage[ps,mover,small,skaknew]{skak}
\usepackage{chessboard}

\makeatletter
\def\notationOn{\let\print@board=\show@board@notation%
  \let\print@inverseboard=\show@board@notation@inverse}
\def\notationOff{\let\print@board=\show@board%
  \let\print@inverseboard=\show@board@inverse}

\renewcommand\showboard{\print@board}
\renewcommand\showinverseboard{\print@inverseboard}
\makeatother

\notationOn

\let\ORIshowboard\showboard
\let\ORIshowinverseboard\showinverseboard
\renewcommand\showboard{%
    \makebox[8\squarelength]{%
    \rule{0pt}{9\squarelength}
    \begin{postscript}
    [trim = \squarelength{} 0pt \squarelength{} 0pt]
    \ORIshowboard
    \end{postscript}}}

\renewcommand\showinverseboard{%
    \makebox[8\squarelength]{%
    \rule{0pt}{9\squarelength}
    \begin{postscript}
    [trim = \squarelength{} 0pt \squarelength{} 0pt]
    \ORIshowinverseboard
    \end{postscript}}}

\newcommand{\safe}[1]{}

\newcommand{\gardnerstart}
           {\fenboard{8/8/1rnbqk2/1ppppp2/8/1PPPPP2/1RNBQK2/8 w - - 0 1}}

\newcommand{\printgardner}
           {\begin{center} \chessboard[printarea=b2-f6] \end{center}}

\newcommand{\varid}[2]
           {$[#1|#2]$}

\newcommand{\bloss}
           {$+ -\ $}

\newcommand{\wloss}
           {$- +\ $}

\newcommand{\draw}
           {$=\ $}

\newcommand{\dtmb}[1]
           {$\sharp #1 \sharp_\bullet$}

\newcommand{\dtmw}[1]
           {$\sharp #1 \sharp_\circ$}

\newcommand{\justif}
           {$\triangle\ $}

\title{Gardner's Minichess Variant is solved}
\author{Mehdi Mhalla \\
Mehdi.Mhalla@imag.fr \\
CNRS Universit\'e de Grenoble - LIG, B.P. 53 \\ 
- 38041 Grenoble Cedex 09, France
\and 
Fr\'ed\'eric Prost \\
Frederic.Prost@imag.fr \\
Universit\'e de Grenoble - LIG, B.P. 53 \\
- 38041 Grenoble Cedex 09, France}

\begin{document}
\maketitle

\begin{abstract}
  A $5 \times 5$ board is the smallest board on which one can set up
  all kind of chess pieces as a start position. We consider Gardner's
  minichess variant in which all pieces are set as in a standard chessboard
  (from Rook to King). This game has roughly $9 \times 10^{18}$ legal
  positions and is comparable in this respect with checkers. We weakly
  solve this game, that is we prove its game-theoretic value and give
  a strategy to draw against best play for White and Black sides.  Our
  approach requires surprisingly small computing power. We give a
  human readable proof. The way the result is obtained is generic and
  could be generalized to bigger chess settings or to other games.
\end{abstract}

\section{Introduction}

  Solving popular games like Othello, Checkers or Chess tantamount to the
grail search in the field of computer games. The resolution of checkers
\cite{SchaBurBjorKis07} put a mark in the field in the sense that
the space search of this game is enormous ($5 \times 10^{20}$) and the
difficulty to make correct move decisions fairly high.

The game of chess have always been recognized as the ultimate
challenge in artificial intelligence. Since the early days of computer
science chess and computers have interacted together
\cite{Pro12}. Nowadays computers have superhuman strength and the game
is partially solved: endgame databases up to few pieces have been
computed.  The most famous ones being the Nalimov tables (6
pieces). Recently Lomonosov endgame tablebases \cite{Lomonosov} have
been computed and give perfect play for 7 pieces (the size of the
tablebase is 140 Terabytes). Nevertheless, the resolution of chess
remains too difficult to be imagined: the number of legal positions is
something around $10^{45}$ \cite{Alis94} and decision complexity is
very high (the amount of chess literature is a proof by itself).

Some studies have been done to resolve particular cases of chess on
smaller board. Notably, $3\times 3$, $3 \times 4$ and $4 \times 4$
(limited to 9 pieces on the board) chess variants have been solved by
K. Kryukov \cite{Kryu33,Kryu34,Kryu44}. In these variants there is
no starting position as in traditional chess. Positions are treated as
puzzles.  Each variant is strongly solved in the sense that the
game-theoretic value of all legal position is determined together with
the perfect play associated. The number of legal positions is
roughly $3 \times 10^{15}$ for the $4\times 4$ variant 
\cite{Kryu44}.

In this paper we study the variant called Gardner's Chess. It is
played on a $5 \times 5$ board, the initial position is the initial
position of chess but for the three pieces on the King side that are
removed. The rules are the ones of classical chess without the two
squares move for Pawns, en passant moves and castling. This variant has
roughly $9 \times 10^{18}$ legal positions. This variant has been
played extensively notably in Italy by correspondence
\cite{Prit07}. The results of finished games were the following:
\begin{itemize}
   \item  White victory 40\%
   \item  draw 32\%
   \item  Black victory 28 \%
\end{itemize}
 
\section{Results}

The game-theoretic value of Gardner's Chess is draw. We prove this by
giving two oracles, one for White and one for Black. Both oracles can
force draw versus best play. The intersection of the two oracles
gives flawless games. Thus Gardner's chess is weakly solved. 

The proof is surprisingly small and can be totally checked by a human.
Oracles are given in appendix \ref{sec:oracle_gardner_White} for
the White side and appendix \ref{sec:oracle_gardner_Black} for the
Black side. From this point of view our result strongly differs from
the resolution of checkers despite the fact that space search and
difficulty of decision are of the same order of magnitude in both
games. Indeed, the proof of \cite{SchaBurBjorKis07} is not
checkable by human eyes: it has required an enormous computing power
(hundreds of computers in parallel over a decade).

Most of our work was achieved with consumer-grade laptop computers. We
have adapted the open source Stockfish chess engine
\cite{Stockfish} to play Gardner's Chess mainly by restricting the
movements to the part of the board and changing the promotion
ranks. 
Sources, executables for several environments and various files,
including the oracles in PGN format as well as the list of the perfect
openings for Gardner's Chess, can be found at the author's Minichess
Resolution page: 
``http://membres-lig.imag.fr/prost/MiniChessResolution/''.

The main line of oracles were computed in a semi-automated way: we
were mainly following the most equalizing line. It turns out that most
of the deviations from the main line can be quickly decided. It is
mainly due to the fact that in Gardner's chess pawns are immediately
exchanged or blocked. Moreover, pieces cannot develop naturally since
almost all free squares are controlled by pawns. Also the fact that
promotion happens quickly leads to some very rapid checkmates that
allow to prune the game tree.

Using these Oracles it is impossible to lose.  Oracle for White
(resp. for Black) does not examine alternative choices for White
(resp. Black) decision nodes but indicates how to answer every
possible Black (resp. White) "reasonable" move. Unreasonable moves,
i.e. moves that obviously lead to a position where it is clear that
Black (resp. White) cannot win can be dealt with our engine. We
provide the maximal number of moves required to mate for our engine
(not necessarily the distance to mate).
Nevertheless, in these positions, from a human point of view, it is
easy not to lose.

 As a by-product of our study on Gardner's Chess the analysis of
perfect openings shows the positions in which the evaluation of
Stockfish is tricked. Indeed for some positions while showing largely
``won'' evaluation (up to +6) the position is completely equal. What is
interesting is that these evaluation bugs can be found on a 8x8 board
as well. Thus the analysis of these positions may help to improve the
evaluation of Stockfish for standard chess games. 

A complete description of the openings in gardner mini chess as
well as a sample of tricky draws and difficult checkmates can be found
at {\em {``http://membres-lig.imag.fr/prost/MiniChessResolution/''}}

\section{Gardner : Oracle for White Draw}
\label{sec:oracle_gardner_White}


We give an oracle for the White side of the Gardner variation. The
objective is to force a draw versus the best play. Therefore, we
give it as a tree of variations that needs no explanations on White nodes:
it is maybe possible to find shorter draw (or even win) but our
aim is to have an oracle the most readable from a human point of
view: the definitive judgment on the leafs of this tree are clear.

Since there are no choices to be explored for White nodes we adopt the
following convention to name sub-variations: first we note the depth in
the oracle, then we enumerate deviations from the main line by
enumerating Black (relevant) moves from left to right, pawns come
first, after we enumerate moves of the pieces following the lexical
order going from left to right and top to bottom. Thus the variation
\varid{3}{1.3.7} is the one obtained by following the oracle until
depth 3 and selecting as sub-variation move 1 as the first move for
Black, then move 3 as second move for Black and 7 as the third
move. We write \bloss (resp. \wloss) when it is obvious that Black
(resp. White) cannot win. We write \dtmb{x} (resp. \dtmw{x}) when
there exists a forced checkmate of the Black King (res. White King) in
$x$ moves (though it is possible that shorter checkmates exist). Very
often positions that look lost (because one side has a piece
advantage for instance) can be fully decided by our engine as forced
checkmates. Justifying lines are written like this: \gardnerstart
\justif \mainline{1.b4 cxb4 2.cxb4 d4}. Finally, the coordinate of the
lower left square is {\bf b2}. Hence the starting position is:

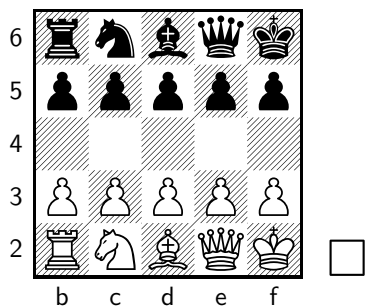
\gardnerstart 
\printgardner

In this position the Black move identified by 1 is \bmove{b4} and move
number 6 is \bmove{Nb4}, move number 7 is \bmove{Nc4}. \bigskip

We give the White oracle as a variation tree. After each
move of the oracle we start by giving all lines in which a forced
checkmate can be found using our engine.

\mainline{1.b4} \storegame{g1} 

 (\color{red}\restoregame{g1}\mainline{1... d4 2.bxc5} \dtmb{47} \justif after
\mainline{2... Bxc5 3.f4} both the pressure on the b file and on
diagonal b2 f6 are too strong to be sustained by by Black. 
\restoregame{g1}\mainline{1...e4 2.bxc5} \dtmb{28}, the point is that
on \mainline{2... Bxc5 3.d4} the threat of \wmove{Rxb5} combined with
the lack of space for Black is too hard to be met.  Other moves just
lose a piece at least.  \restoregame{g1}\mainline{1...f4 2.bxc5}
\dtmb{24} \justif{2... Bxc5 3.d4} the threat of \wmove{Rxb5}.
\restoregame{g1}\mainline{1...Nxb4 2.cxb4} \dtmb{24} White is a piece
up for nothing.  \restoregame{g1}\mainline{1...Nd4 2.bxc5} \dtmb{17}
White is a piece up for nothing.\color{black}):

\begin{itemize}
  \item \varid{1}{1} \restoregame{g1}\mainline{1... c4 2.d4} \storegame{g11}
         (\color{red}\restoregame{g11}\mainline{2... Bxb4 3.dxe5+} \dtmb{29},
          \restoregame{g11}\mainline{2... Bc5 3.bxc5} \dtmb{8},
          \restoregame{g11}\mainline{2... Nxd4} \dtmb{29},
          \restoregame{g11}\mainline{2... f4 3.e4} \dtmb{38}\color{black}). 
  \begin{itemize}
  \item \varid{1}{1.1} \restoregame{g11}\mainline{2... exd4 3.exd4}
     \storegame{g111}  
     (\color{red}\restoregame{g111}\mainline{3... Bc5} \dtmb{6},
      \restoregame{g111}\mainline{3... Ne5} \dtmb{14},
      \restoregame{g111}\mainline{3... Qe3+} \dtmb{6},
      \restoregame{g111}\mainline{3... Qe4} \dtmb{10},
      \restoregame{g111}\mainline{3... Qe5} \dtmb{12},
      \restoregame{g111}\mainline{3... Be5}\dtmb{31}\color{black}):

    \begin{itemize}
    \item \varid{1}{1.1.1} \restoregame{g111}\mainline{3... f4
        4.Qxe6+ Kxe6} \draw White just has to move his King on e2-f2
      and Black cannot break through. No matter what is the relative
      position of the two Kings, if black Knight takes on d4 or b4
      White takes back with the Knight and the position is still
      blocked for Black and if Black plays \bmove{Bxb4} the position
      is \dtmb{17} when kings on file e and \dtmb{24} when kings are
      in file  f. Finally, if Black plays \bmove{Ne5} white just 
      takes it with \wmove{dxe5} and if Black plays \bmove{Be5}
      White just continue to move his king.

      \item  \varid{1}{1.1.2} \restoregame{g111}\mainline{3... Qxe2+
          4.Kxe2} \draw for the same reason as line \varid{1}{1.1.1}.
    \end{itemize} 

    \item \varid{1}{1.2} \restoregame{g11}\mainline{2... e4 3.f4} \draw
    Black is in zugzwang and must give a piece. Due to the blocked
    nature of the position he can do it without losing but he cannot
    break through e.g. \mainline{3... Be5 4.fxe5+ Nxe5 5.dxe5+ Qxe5
      6.Nd4} and White can simply moves back and forth with the
    Knight.

    \item \varid{1}{1.3} \restoregame{g11}\mainline{2... Nxb4 3.dxe5+}
    \storegame{g114}
    (\color{red}  \restoregame{g114}\mainline{3... Bxe5 4.Rxb4}
          \dtmb{20},           
           \restoregame{g114}\mainline{3... Kxe5 4.cxd4} \dtmb{13}\color{black})
    \bloss: 
   
      \begin{itemize}
     
        \item \varid{1}{1.3.1} \restoregame{g114}\mainline{3... Qxe5
            4.Nxb4} \storegame{g1142} 
          (\color{red}\restoregame{g1142}\mainline{4... Bxb4 5.Rxb4} \dtmb{25},
          \restoregame{g1142}\mainline{4... Qe6 5.e4} \dtmb{17},
          \restoregame{g1142}\mainline{4... Qd4 5.exd4} \dtmb{3},
          \restoregame{g1142}\mainline{4... Qe4 5.fxe4} \dtmb{9},
          \restoregame{g1142}\mainline{4... Qf4 5.exf4} \dtmb{4},
          \restoregame{g1142}\mainline{4... Qxd3 5.exf4} \dtmb{4},
          \restoregame{g1142}\mainline{4... Qxb3 5.Bxb3+} \dtmb{5},
          \restoregame{g1142}\mainline{4... d4 5.cxd4} \dtmb{10}, 
          \restoregame{g1142}\mainline{4... f4 5.exf4} \dtmb{10},
          \restoregame{g1142}\mainline{4... Rc6 5.Nxc6} \dtmb{10}\color{black}):

           - \varid{1}{1.3.1.1} \restoregame{g1142}
            \mainline{4... Bc5 5.Nc2} \draw White 
             blocks the position on the dark squares
             with \wmove{Nd4} and \wmove{Rb4} (and moves 
             his Rook between b2-b4 if Black does not move.
             \storegame{g11424} \justif \mainline{5... f4 
             6.Nd4 b4} (other moves leads to a loss for Black)
             \mainline{7.exf4 Bxd4+ 8.cxd4 Qxe2+} (other moves lead
             to direct checkmate).

           - \varid{1}{1.3.1.2} \restoregame{g1142}
            \mainline{4... Ke6 5.Nc2} \bloss similar to line 
            \varid{1}{1.3.1.1}.     
      \end{itemize} 
  \end{itemize}
\end{itemize}

\restoregame{g1}\mainline{1... cxb4 2.cxb4} \storegame{g2}
 All Knight and Bishop moves lose a piece and end up in a position
 where clearly Black cannot win
(\color{red}\restoregame{g2}\mainline{2... Nd4} \dtmb{23}, 
  \restoregame{g2}\mainline{2... Nxb4} \dtmb{18},
 \restoregame{g2}\mainline{2... Bxb4} \dtmb{15},  
 \restoregame{g2}\mainline{2... e4 3.Bc3+} \dtmb{15},
 \restoregame{g2}\mainline{2... f4 3.bxc5} \dtmb{29} the idea is 
      that the b6 pawn is lost and Black is lacking space 
      \justif \mainline{3... Bxc5 4.d4 Bd65.Rxb5}\color{black})

\restoregame{g2} 
\mainline{2... d4 3.e4} \storegame{g3}
 (\color{red}\restoregame{g3}\mainline{3... fxe4 4.fxe4} \dtmb{8} 
     since the threat of \wmove{Qf3+} is too strong, 
  \restoregame{g3}\mainline{3... Nxb4} \dtmb{15},
  \restoregame{g3}\mainline{3... Bxb4} \dtmb{15},  
  \restoregame{g3}\mainline{3... Qb3}   \dtmb{12},  
  \restoregame{g3}\mainline{3... Qc4}   \dtmb{8}, 
  \restoregame{g3}\mainline{3... Qc6}\dtmb{8}\color{black})

\restoregame{g3}
\mainline{3... f4 4.Bxf4}
\storegame{g4} (\color{red}All Black's alternatives 
lead to forced checkmate since they lose a piece for nothing 
  \restoregame{g4}\mainline{4... Bc5}  \dtmb{8}, 
  \restoregame{g4}\mainline{4... Qb3} \dtmb{11}, 
  \restoregame{g4}\mainline{4... Qc4} \dtmb{8},  
  \restoregame{g4}\mainline{4... Qd5} \dtmb{5}, 
  \restoregame{g4}\mainline{4... Qf5} \dtmb{8},
  \restoregame{g4}\mainline{4... Nxb4} \dtmb{25},
  \restoregame{g4}\mainline{4... Bxb4} \dtmb{25}\color{black})

\restoregame{g4}
\mainline{4... exf4 5.Qd2}
\storegame{g5} 
(\color{red}\restoregame{g5}\mainline{5... Ne5 6.Qxf4+} \dtmb{2}, 
 \restoregame{g5}\mainline{5... Nxb4 6.Nxb4}
  \bloss d5 cannot be protected and the threat of \wmove{Qxf4+}
  forbids \mainline{6... Bxb4}\color{black})

\restoregame{g5}
\mainline{5... Be5 6.Ke2} \draw White just moves his King on e2-f2 and Black
cannot untangle by \bmove{Ne5} because of \wmove{Qxf4} and must
otherwise give a piece and cannot win. 

\section{Gardner : Oracles for Black Draw}
   \label{sec:oracle_gardner_Black}
  
   We give, for each of the White seven legal first move, an oracle
   from the Black point of view that forces the draw. So here we give
   no explanations for Black decision nodes and we explore all
   reasonable moves (as explained erlier) at White decision
   nodes. These oracles are sometimes much simpler than the White
   oracle for draw since, rather curiously, it is sometimes more
   difficult for White to achieve draw. It means that often even
   slight deviations from the main line directly lead to positions
   that can be decided as forced checkmates.

   \subsection{White moves \wmove{b4}}

\gardnerstart
\mainline{1.b4 cxb4} \storegame{gba1}
(\color{red}\restoregame{gba1}\mainline{2.Rb3 d4} \dtmw{17},
 \restoregame{gba1}\mainline{2.Rxb4 Nxb4} \dtmw{21},
 \restoregame{gba1}\mainline{2.c4 bxc4} \dtmw{15} the b4 c4 pawn 
 duo is too strong, \restoregame{gba1}\mainline{2.Nd4 bxc3} \dtmw{20},
 \restoregame{gba1} \mainline{2.e4 bxc3} \dtmw{25}\color{black}) 
   \begin{itemize}

     \item \varid{1}{1} \restoregame{gba1}\mainline{2.d4 bxc3} 
           \storegame{gba12}
           (\color{red}\restoregame{gba12}\mainline{3.dxe5+} \dtmw{17},
            \restoregame{gba12}\mainline{3.e4} \dtmw{9},
            \restoregame{gba12}\mainline{3.f4} \dtmw{10},
            \restoregame{gba12}\mainline{3.Rb3} \dtmw{17},
            \restoregame{gba12}\mainline{3.Rb4} \dtmw{12},
            \restoregame{gba12}\mainline{3.Rxb5} \dtmw{12},
            \restoregame{gba12}\mainline{3.Na4} \dtmw{9},
            \restoregame{gba12}\mainline{3.Qd3} \dtmw{10},
            \restoregame{gba12}\mainline{3.Qc4} \dtmw{10},
            \restoregame{gba12}\mainline{3.Qxb5} \dtmw{7}\color{black})

            \restoregame{gba12}\mainline{3.Bxc3 b4}
             \storegame{gba128}
              (\color{red}\restoregame{gba128}\mainline{4.e4 dxe4} \dtmw{19}, 
               \restoregame{gba128}\mainline{4.f4 exf4} \dtmw{30},
               \restoregame{gba128}\mainline{4.Nxb4 Nxb4} \dtmw{26}, 
               \restoregame{gba128}\mainline{4.Bxb4 Nxb4} \dtmw{24}, 
               \restoregame{gba128}\mainline{4.Bd2 b3} \dtmw{23}, 
               \restoregame{gba128}\mainline{4.Qd2 bxc3} \dtmw{18}, 
               \restoregame{gba128}\mainline{4.Qd3 e4} \dtmw{26}, 
               \restoregame{gba128}\mainline{4.Qc4 dxc4} \dtmw{10},
               \restoregame{gba128}\mainline{4.Qb5 Rxb5} \dtmw{9}\color{black})
      \begin{itemize}
         \item \varid{1}{1.1} \restoregame{gba128}\mainline{4.dxe5+
           Bxe5} \storegame{gba1281} 
         (\color{red}\restoregame{gba1281}\mainline{5.e4} \dtmw{13},
         \restoregame{gba1281}\mainline{5.f4} \dtmw{15},
         \restoregame{gba1281}\mainline{5.Rb3} \dtmw{15},
         \restoregame{gba1281}\mainline{5.Nxb4} \dtmw{22},
         \restoregame{gba1281}\mainline{5.Nd4} \dtmw{10},
         \restoregame{gba1281}\mainline{5.Qd2} \dtmw{8},
         \restoregame{gba1281}\mainline{5.Qd3} \dtmw{15},
         \restoregame{gba1281}\mainline{5.Qc4} \dtmw{10},
         \restoregame{gba1281}\mainline{5.Qb5} \dtmw{8},
         \restoregame{gba1281}\mainline{5.Bxb4} \dtmw{17},
         \restoregame{gba1281}\mainline{5.Bd2} \dtmw{11},
         \restoregame{gba1281}\mainline{5.Bd4} \dtmw{29},  
         \restoregame{gba1281}\mainline{5.Bxe5+ Qxe5} \dtmw{39}
         \storegame{gba1281a}\color{black})

         \restoregame{gba1281}\mainline{5.Rxb4 Rxb4} 
           \storegame{gba1281ap} 

         (\color{red}\restoregame{gba1281ap}\mainline{6.e4} \dtmw{9},
         \restoregame{gba1281ap}\mainline{6.f4} \dtmw{9},
         \restoregame{gba1281ap}\mainline{6.Bb2} \dtmw{8},
         \restoregame{gba1281ap}\mainline{6.Bd2} \dtmw{10},
         \restoregame{gba1281ap}\mainline{6.Bd4} \dtmw{9},
         \restoregame{gba1281ap}\mainline{6.Bxe5+} \dtmw{51?},
         \restoregame{gba1281ap}\mainline{6.Nxb4} \dtmw{27},
         \restoregame{gba1281ap}\mainline{6.Nd4} \dtmw{12},
         \restoregame{gba1281ap}\mainline{6.Qd2} \dtmw{11},
         \restoregame{gba1281ap}\mainline{6.Qd3} \dtmw{20},
         \restoregame{gba1281ap}\mainline{6.Qc4} \dtmw{6},
         \restoregame{gba1281ap}\mainline{6.Qb5} \dtmw{7}\color{black})

        \restoregame{gba1281ap} \mainline{6.Bxb4 Nxb4}
        \storegame{gba1281apa}
        (\color{red}\restoregame{gba1281apa} \mainline{7.e4}\dtmw{7},
        \restoregame{gba1281apa} \mainline{7.Nd4}\dtmw{14},
        \restoregame{gba1281apa} \mainline{7.Qb5}\dtmw{9},
        \restoregame{gba1281apa} \mainline{7.Qc4}\dtmw{6},
        \restoregame{gba1281apa} \mainline{7.Qd3}\dtmw{6},
        \restoregame{gba1281apa} \mainline{7.Qd2}\dtmw{17}\color{black})
        \begin{itemize}
          \item \varid{1}{1.1.1}\restoregame{gba1281apa} \mainline{7.f4 Bc3}
                 \storegame{gba1281apb}
             (\color{red}\restoregame{gba1281apb} \mainline{8.e4} \dtmw{8},
             \restoregame{gba1281apb} \mainline{8.Nd4} \dtmw{13},
             \restoregame{gba1281apb} \mainline{8.Qb5} \dtmw{10},
             \restoregame{gba1281apb} \mainline{8.Qc4} \dtmw{5},
             \restoregame{gba1281apb} \mainline{8.Qd3} \dtmw{5},
             \restoregame{gba1281apb} \mainline{8.Qd2} \dtmw{4},
             \restoregame{gba1281apb} \mainline{8.Qe2} \dtmw{8},
             \restoregame{gba1281apb} \mainline{8.Ke2} \dtmw{7}\color{black})
             \restoregame{gba1281apb} \mainline{8.Nxb4 d4} 
             \storegame{gba1281apc}
             (\color{red}\restoregame{gba1281apc}\mainline{9.exd4} \dtmw{18},
             \restoregame{gba1281apc}\mainline{9.Nd5} \dtmw{17},
             \restoregame{gba1281apc}\mainline{9.Qd3} \dtmw{16},
             \restoregame{gba1281apc}\mainline{9.Qf3} \dtmw{16},
             \restoregame{gba1281apc}\mainline{9.Nc6} \dtmw{14},
             \restoregame{gba1281apc}\mainline{9.Kf3} \dtmw{10},
             \restoregame{gba1281apc}\mainline{9.e4} \dtmw{7},
             \restoregame{gba1281apc}\mainline{9.Qd2} \dtmw{5},
             \restoregame{gba1281apc}\mainline{9.Qb2 Qxe3} checkmate,
             \restoregame{gba1281apc}\mainline{9.Qc2 Qxe3} checkmate,
             \restoregame{gba1281apc}\mainline{9.Qc4 Qxe3} checkmate,
             \restoregame{gba1281apc}\mainline{9.Qb5 Qxe3} checkmate\color{black}) 
             \draw
             since on both \wmove{Nc2} and \wmove{Nd3} black exchanges 
             Queen on e3 and the remaining position is draw \justif
            \restoregame{gba1281apc} \mainline{9.Nc2 dxe3+ 10.Nxe3 Bd4 11.Qe2 Bxe3+ 12.Qxe3
              Qxe3+ 13. Kxe3}.

        \item \varid{1}{1.1.2} \restoregame{gba1281apa} 
               \mainline{7.Nxb4 d4} \draw the only move to 
              avoid \bmove{dxe3+} and the liquidation of all pawns 
              is \mainline{8.e3 Qb3 9.Nd5+ Ke6 10.exf5+ Kxd5
              11.f4 Qe3+ 12.Qxe3 dxe3+ 13.Kxe3}. 
        \end{itemize}
    
       \safe{
         (\color{red}\restoregame{gba1281a}\mainline{6.e4} \dtmw{10},
         \restoregame{gba1281a}\mainline{6.f4} \dtmw{8},
         \restoregame{gba1281a}\mainline{6.Rxb4} \dtmw{11},
         \restoregame{gba1281a}\mainline{6.Nxb4} \dtmw{22},
         \restoregame{gba1281a}\mainline{6.Nd4} \dtmw{19},
         \restoregame{gba1281a}\mainline{6.Qd2} \dtmw{10},
         \restoregame{gba1281a}\mainline{6.Qb5} \dtmw{7},
         \restoregame{gba1281a}\mainline{6.Qc4} \dtmw{6},
         \restoregame{gba1281a}\mainline{6.Qd3} \dtmw{7}\color{black}) 

         \restoregame{gba1281a}\mainline{6.Rb3 f4}
          \storegame{gba1281b}
        (\color{red}\restoregame{gba1281b}\mainline{7.e4} \dtmw{16}, 
        \restoregame{gba1281b}\mainline{7.exf4} \dtmw{37}, 
        \restoregame{gba1281b}\mainline{7.Rb2} \dtmw{6}, 
        \restoregame{gba1281b}\mainline{7.Rxb4} \dtmw{16}, 
        \restoregame{gba1281b}\mainline{7.Rc3} \dtmw{6}, 
        \restoregame{gba1281b}\mainline{7.Rd3} \dtmw{9}, 
        \restoregame{gba1281b}\mainline{7.Nxb4} \dtmw{13}, 
        \restoregame{gba1281b}\mainline{7.Nd4} \dtmw{8}, 
        \restoregame{gba1281b}\mainline{7.Qd3} \dtmw{29}, 
        \restoregame{gba1281b}\mainline{7.Qc4} \dtmw{5}, 
        \restoregame{gba1281b}\mainline{7.Qb5} \dtmw{8}\color{black}) 

        \restoregame{gba1281b}\mainline{7.Qd2} MAT EN 37}

         \item \varid{1}{1.2} \restoregame{gba128}\mainline{4.Rb3
             f4} \storegame{gba1284}
             (\color{red}\restoregame{gba1284}\mainline{5.e4} \dtmw{19},  
             \restoregame{gba1284}\mainline{5.Rb2} \dtmw{33}, 
             \restoregame{gba1284}\mainline{5.Rxb4} \dtmw{27}, 
             \restoregame{gba1284}\mainline{5.Bxb4} \dtmw{22}, 
             \restoregame{gba1284}\mainline{5.Nxb4} \dtmw{27}, 
             \restoregame{gba1284}\mainline{5.Qb5} \dtmw{10},
             \restoregame{gba1284}\mainline{5.Qc4} \dtmw{8},
             \restoregame{gba1284}\mainline{5.Qd2} \dtmw{20}\color{black})
             \begin{itemize}
               \item \varid{1}{1.2.1} \restoregame{gba1284} 
                     \mainline{5.dxe5+ Bxe5} \storegame{gba1284a}
             (\color{red}\restoregame{gba1284a}\mainline{6.e4} \dtmw{9},
              \restoregame{gba1284a}\mainline{6.exf4} \dtmw{17},
              \restoregame{gba1284a}\mainline{6.Rb2} \dtmw{11},
              \restoregame{gba1284a}\mainline{6.Rxb4} \dtmw{32},
              \restoregame{gba1284a}\mainline{6.Bxb4} \dtmw{39},
              \restoregame{gba1284a}\mainline{6.Bb2} \dtmw{22},
              \restoregame{gba1284a}\mainline{6.Bd2} \dtmw{21},
              \restoregame{gba1284a}\mainline{6.Bxe5+} \dtmw{60},
              \restoregame{gba1284a}\mainline{6.Nd4} \dtmw{9},
              \restoregame{gba1284a}\mainline{6.Qd2} \dtmw{14},
              \restoregame{gba1284a}\mainline{6.Qd3} \dtmw{12},
              \restoregame{gba1284a}\mainline{6.Qc4} \dtmw{7},
              \restoregame{gba1284a}\mainline{6.Qb5} \dtmw{10},
              \restoregame{gba1284a}\mainline{6.Nxb4} \dtmw{20}\color{black})

              \restoregame{gba1284a}\mainline{6.Bd4 Nxd4} 
              \storegame{gba1284b} 
              (\color{red}\restoregame{gba1284b}\mainline{7.Nxd4} \dtmw{25},
                 \restoregame{gba1284b}\mainline{7.Qd2} \dtmw{9},
                 \restoregame{gba1284b}\mainline{7.Qd3} \dtmw{9},
                 \restoregame{gba1284b}\mainline{7.Nxb4} \dtmw{9},
                 \restoregame{gba1284b}\mainline{7.exf4} \dtmw{8},
                 \restoregame{gba1284b}\mainline{7.Rxb4} \dtmw{7},
                 \restoregame{gba1284b}\mainline{7.Rb2} \dtmw{7},
                 \restoregame{gba1284b}\mainline{7.Rd3} \dtmw{7},
                 \restoregame{gba1284b}\mainline{7.e4} \dtmw{7},
                 \restoregame{gba1284b}\mainline{7.Qb5} \dtmw{7},
                 \restoregame{gba1284b}\mainline{7.Rc3} \dtmw{6},
                 \restoregame{gba1284b}\mainline{7.Qc4} \dtmw{6}\color{black})

               \restoregame{gba1284b}\mainline{7.exd4 Bd6} 
               \storegame{gba1284c} 
               (\color{red}\restoregame{gba1284c}\mainline{8.Qd3} \dtmw{26},
               \restoregame{gba1284c}\mainline{8.Rb2} \dtmw{23},
               \restoregame{gba1284c}\mainline{8.Qd2} \dtmw{22}, 
               \restoregame{gba1284c}\mainline{8.Nxb4} \dtmw{18}, 
               \restoregame{gba1284c}\mainline{8.Rxb4} \dtmw{13},
               \restoregame{gba1284c}\mainline{8.Ne3} \dtmw{11}, 
               \restoregame{gba1284c}\mainline{8.Rc3} \dtmw{10},
               \restoregame{gba1284c}\mainline{8.Qe3} \dtmw{7},
               \restoregame{gba1284c}\mainline{8.Qe4} \dtmw{6},
               \restoregame{gba1284c}\mainline{8.Qe5} \dtmw{7},
               \restoregame{gba1284c}\mainline{8.Rd3} \dtmw{10},
               \restoregame{gba1284c}\mainline{8.Re3} \dtmw{8},
               \restoregame{gba1284c}\mainline{8.Qb5} \dtmw{6}\color{black})
     
               \restoregame{gba1284c} 
               \mainline{8.Qxe6+ Kxe6} 
               \draw White is blocked by Black pawns and cannot 
               progress. The Black King may just move on f6 f5 squares. 

               \item \varid{1}{1.2.2}\restoregame{gba1284}
                 \mainline{5.exf4 exf4} \storegame{gba12841} \draw 
                 \justif \mainline{6.Qd3 Qf5 7.Qe2 Qe6} with
                 repetition. White cannot play \wmove{Qd2} due to 
                 \bmove{bxc3}. Once queen have been exchanged the
                 position is blocked and Black can just move his
                 King ad lib.

               \item \varid{1}{1.2.3}\restoregame{gba1284}
                 \mainline{5.Bb2 fxe3+ 6.Qxe3 exd4 7.Bxd4+ exd4
                 8.Qxd4+ Be5} \draw Black can play \bmove{Bc3} on any
               queen moves and blocks the position. 

               \item \varid{1}{1.2.4}\restoregame{gba1284}
                 \mainline{5.Bd2 exd4 6.exd4 Qxe2+ 7.Kxe2} \draw
                 White cannot untangle and Black may just move his
                 King around. 

               \item \varid{1}{1.2.5}\restoregame{gba1284}
                 \mainline{5.Qd3 exd4} \draw similar as line 
                 \varid{1}{2.8.4.1}.
             \end{itemize}

         \item \varid{1}{1.3} \restoregame{gba128}\mainline{4.Rxb4 
               Rxb4} \draw \justif \mainline{5.dxe5+ Bxe5 6.Bxb4
               Nxb4 7.Nxb4 d4}.
    \end{itemize} 

     \item \varid{1}{2} \restoregame{gba1}\mainline{2.f4 bxc3} 
        \storegame{gba14}
        (\color{red}\restoregame{gba14}\mainline{3.d4} \dtmw{10},
         \restoregame{gba14}\mainline{3.e4} \dtmw{9},
         \restoregame{gba14}\mainline{3.fxe5+ Bxe5} \dtmw{15}, 
         \restoregame{gba14}\mainline{3.Rb3} \dtmw{15},
         \restoregame{gba14}\mainline{3.Rb4} \dtmw{15},
         \restoregame{gba14}\mainline{3.Rxb5} \dtmw{14},
         \restoregame{gba14}\mainline{3.Nb4} \dtmw{10},
         \restoregame{gba14}\mainline{3.Nd4} \dtmw{14},
         \restoregame{gba14}\mainline{3.Qf3} \dtmw{8},
         \restoregame{gba14}\mainline{3.Kf3} \dtmw{8}\color{black})

         \restoregame{gba14}\mainline{3.Bxc3 b4} \storegame{gba149}
          (\color{red}\restoregame{gba149}\mainline{4.d4} \dtmw{33}, 
           \restoregame{gba149}\mainline{4.e4} \dtmw{12}, 
           \restoregame{gba149}\mainline{4.Bxb4} \dtmw{26}, 
           \restoregame{gba149}\mainline{4.Bd4} \dtmw{22},
           \restoregame{gba149}\mainline{4.Bd2 exf4} \dtmw{23},
           \restoregame{gba149}\mainline{4.Kf3} \dtmw{22},
           \restoregame{gba149}\mainline{4.Qd2} \dtmw{22},
           \restoregame{gba149}\mainline{4.Nd4} \dtmw{19}\color{black}) 

       \begin{itemize}
         \item \varid{1}{2.1} \restoregame{gba149}
           \mainline{4.fxe5+ Bxe5} \storegame{gba1493}
                 (\color{red}\restoregame{gba1493}\mainline{5.e4} \dtmw{12},
                  \restoregame{gba1493}\mainline{5.Rb3} \dtmw{19},
                  \restoregame{gba1493}\mainline{5.Rxb4} \dtmw{27},
                  \restoregame{gba1493}\mainline{5.Nxb4} \dtmw{26},
                  \restoregame{gba1493}\mainline{5.Nd4} \dtmw{10},
                  \restoregame{gba1493}\mainline{5.Bxb4} \dtmw{14},
                  \restoregame{gba1493}\mainline{5.Bd4} \dtmw{23},
                  \restoregame{gba1493}\mainline{5.Bd2} \dtmw{11},
                  \restoregame{gba1493}\mainline{5.Qd2} \dtmw{14},
                  \restoregame{gba1493}\mainline{5.Qf3} \dtmw{12},
                  \restoregame{gba1493}\mainline{5.Kf3} \dtmw{11}\color{black})          
            \begin{itemize}
            \item  \varid{1}{2.1.1} \restoregame{gba1493}
            \mainline{5.Bxe5+ Qxe5} \storegame{gba14931} 
            (\color{red}\restoregame{gba14931}\mainline{6.e4} \dtmw{10}, 
            \restoregame{gba14931}\mainline{6.Rxb4} \dtmw{12}, 
            \restoregame{gba14931}\mainline{6.Nxb4} \dtmw{15}, 
            \restoregame{gba14931}\mainline{6.Nd4} \dtmw{14}, 
            \restoregame{gba14931}\mainline{6.Qd2} \dtmw{8}, 
            \restoregame{gba14931}\mainline{6.Qf3} \dtmw{8}, 
            \restoregame{gba14931}\mainline{6.Kf3} \dtmw{8}\color{black})
            
             - \varid{1}{2.1.1.1}\restoregame{gba14931}
             \mainline{6.Rb3 d4} \storegame{gba149311} 
              (\color{red}\restoregame{gba149311}\mainline{7.Qf3} \dtmw{20},
                 \restoregame{gba149311}\mainline{7.exd4} \dtmw{17},
                 \restoregame{gba149311}\mainline{7.Qd2} \dtmw{14},
                 \restoregame{gba149311}\mainline{7.e4} \dtmw{13},
                 \restoregame{gba149311}\mainline{7.Kf3} \dtmw{12},
                 \restoregame{gba149311}\mainline{7.Rb2} \dtmw{12},
                 \restoregame{gba149311}\mainline{7.Nxb4} \dtmw{11},
                 \restoregame{gba149311}\mainline{7.Rxb4} \dtmw{11},
                 \restoregame{gba149311}\mainline{7.Rc3} \dtmw{8}\color{black})

               \restoregame{gba149311} \mainline {7.Nxd4 Nxd4}              
               \storegame{gba1493111} 
               (\color{red}\restoregame{gba1493111}\mainline{8.Qb2} \dtmw{9},
                 \restoregame{gba1493111}\mainline{8.Rxb4} \dtmw{8},
                 \restoregame{gba1493111}\mainline{8.Qd2} \dtmw{7},
                 \restoregame{gba1493111}\mainline{8.Rb2} \dtmw{7},
                 \restoregame{gba1493111}\mainline{8.Qc2} \dtmw{5},
                 \restoregame{gba1493111}\mainline{8.Rc3} \dtmw{4},
                 \restoregame{gba1493111}\mainline{8.e4} \dtmw{2}\color{black})

                 \restoregame{gba1493111} \mainline {8.exd4 Qxd4+}      
                  \storegame{gba14931111} 
                 (\color{red}\restoregame{gba14931111}\mainline{9.Kf3} \dtmw{12}\color{black})
      
                 \restoregame{gba14931111} \mainline{9.Qe3 Qxe3+ 
                 10.Kxe3 Ke5} \wloss Black can easily achieve draw
               since the White Rook has to keep an eye on the b pawn
               and the Black King is in front of the White d pawn.

             - \varid{1}{2.1.1.2} \restoregame{gba14931}
             \mainline{6.d4 Qe4 7.Rb3 f4} \draw since the only moves
             that not lose for White are \storegame{gba14931} either 
             \mainline{8.exf4 Qxe2+ 9. Kxe2 Kf5} and Black and White
             King move ad lib., or \restoregame{gba14931}
             \mainline{8.Qd2 f3 9.Nxb4 Rxb4 10.Rxb4 Nxb4 11.Qxb4 Qc2+}
             and perpetual check or \mainline{12.Qc5 Qb3 13.Qd6 Qxe3+
               14.Kxe3} stalemate.  
 
            \item  \varid{1}{2.1.2} \restoregame{gba1493} 
            \mainline{5.d4 Bd6} \storegame{gba1412b}
            (\color{red}\restoregame{gba1412b}\mainline{6.e4} \dtmw{15},
             \restoregame{gba1412b} \mainline{6.Nxb4} \dtmw{26},
             \restoregame{gba1412b} \mainline{6.Bxb4} \dtmw{30},
             \restoregame{gba1412b} \mainline{6.Bd2} \dtmw{45},
             \restoregame{gba1412b} \mainline{6.Rxb4} \dtmw{14},
             \restoregame{gba1412b} \mainline{6.Qd2} \dtmw{21},
             \restoregame{gba1412b} \mainline{6.Qd3} \dtmw{24},
             \restoregame{gba1412b} \mainline{6.Qc4} \dtmw{18},
             \restoregame{gba1412b} \mainline{6.Qb5} \dtmw{8},
             \restoregame{gba1412b} \mainline{6.Qf3} \dtmw{32},
             \restoregame{gba1412b} \mainline{6.Kf3} \dtmw{16}\color{black}) 

            \restoregame{gba1412b}
            \mainline{6.Rb3 Qe4}
            \wloss since White cannot do anything to untangle and Black
            may just move his Rook on b5 b6. 
            \end{itemize}

         \item \varid{1}{2.2} \restoregame{gba149}\mainline{4.Rb3
             d4} \storegame{gba1494} 
              (\color{red}\restoregame{gba1494}\mainline{5.e4} \dtmw{10},
              \restoregame{gba1494}\mainline{5.Rb2} \dtmw{10},
              \restoregame{gba1494}\mainline{5.Rxb4} \dtmw{20},
              \restoregame{gba1494}\mainline{5.Nxb4} \dtmw{13},
              \restoregame{gba1494}\mainline{5.Nxd4} \dtmw{14},
              \restoregame{gba1494}\mainline{5.Bb2} \dtmw{14},
              \restoregame{gba1494}\mainline{5.Bxb4} \dtmw{15},
              \restoregame{gba1494}\mainline{5.Bxd4} \dtmw{14},
              \restoregame{gba1494}\mainline{5.Bd2} \dtmw{11},
              \restoregame{gba1494}\mainline{5.Qd2} \dtmw{11},
              \restoregame{gba1494}\mainline{5.Qf3} \dtmw{10},
              \restoregame{gba1494}\mainline{5.Kf3} \dtmw{9}\color{black})
         \begin{itemize}
           \item \varid{1}{2.2.1}  \restoregame{gba1494}
                  \mainline{5.exd4 exd4 6.Qxe6+}, other White moves
                  lose straightforwardly since the Bisop is lost, 
                  \mainline{6... Kxe6 7.Bxd4}, otherwise Black just
                  moves his Rook on b5 b6 and White cannot break
                  through \mainline{7... Nxd4 8.Nxd4+ Kd5}  \draw the
                  f4 pawn is going to fall and White cannot win this position.
           \item \varid{1}{2.2.2}  \restoregame{gba1494}
                  \mainline{5.fxe5+ Bxe5 6.Rxb4} other moves lose the
                  Rook and lead to quick White defeat
                  \mainline{6... Rxb4 7.Bxb4 Nxb4 8.Nxb4 dxe3+ 9.Qxe3
                    Qd6 10.Nc2 Qc6} \draw the best for White is to
                  repeat moves with \mainline{11.Nb4 Qd6}.   
         \end{itemize}

         \item \varid{1}{2.3} \restoregame{gba149}\mainline{4.Rxb4
             Rxb4} \storegame{gba1495} 
              (\color{red}\restoregame{gba1495}\mainline{5.d4} \dtmw{20},
              \restoregame{gba1495}\mainline{5.e4} \dtmw{12},
              \restoregame{gba1495}\mainline{5.Nd4} \dtmw{12}, 
              \restoregame{gba1495}\mainline{5.Bxb4} \dtmw{20},
              \restoregame{gba1495}\mainline{5.Bb2} \dtmw{9},
              \restoregame{gba1495}\mainline{5.Bd4} \dtmw{14},
              \restoregame{gba1495}\mainline{5.Bd2} \dtmw{11}, 
              \restoregame{gba1495}\mainline{5.Bxe5+} \dtmw{18},
              \restoregame{gba1495}\mainline{5.Qd2} \dtmw{16},
              \restoregame{gba1495}\mainline{5.Qf3} \dtmw{12},
              \restoregame{gba1495}\mainline{5.Kf3} \dtmw{12}\color{black})
         \begin{itemize}
           \item \varid{1}{2.3.1}  \restoregame{gba1495}
                  \mainline{5.fxe5+ Bxe5 6.Bxb4 Nxb4 7.Nxb4 d4} 
                 \draw last pawns will soon be exchanged and White
                 cannot force any advantage \justif \mainline{8.e4 Qb3
                 9.Nd5+ Ke6 10.Qf3 fxe4}.

           \item \varid{1}{2.3.2}  \restoregame{gba1495}
                  \mainline{5.Nxb4 Nxb4 6.fxe5+ Bxe5 7.Bxb4 d4} \draw 
                 for the same reasons as line \varid{1}{4.3.1}.
         \end{itemize}             

         \item \varid{1}{2.4} \restoregame{gba149}\mainline{4.Bxe5+
           Bxe5} \storegame{gba1498}
         (\color{red}\restoregame{gba1498}\mainline{5.d4} \dtmw{24},
         \restoregame{gba1498}\mainline{5.e4} \dtmw{10},
         \restoregame{gba1498}\mainline{5.Rb3} \dtmw{21},
         \restoregame{gba1498}\mainline{5.Rxb4} \dtmw{13},
         \restoregame{gba1498}\mainline{5.Nxb4} \dtmw{14},
         \restoregame{gba1498}\mainline{5.Nd4} \dtmw{14},
         \restoregame{gba1498}\mainline{5.Qd2} \dtmw{9},
         \restoregame{gba1498}\mainline{5.Qf3} \dtmw{9},
         \restoregame{gba1498}\mainline{5.Kf3} \dtmw{9}\color{black})
         
         \restoregame{gba1498}\mainline{5.fxe5+ Qxe5 6.d4 Qe4}
         \draw the position is blocked and Black can just move his
         King to e6 f6 White can't remove his Rook from the b file and
         if he tries to break through the ending will be a clear draw.

         \item \varid{1}{2.5} \restoregame{gba149}\mainline{4.Nxb4
           Rxb4} \storegame{gba14910}
               (\color{red}\restoregame{gba14910}\mainline{5.d4} \dtmw{20},
                \restoregame{gba14910}\mainline{5.e4} \dtmw{12},
                \restoregame{gba14910}\mainline{5.Rb3} \dtmw{10},
                \restoregame{gba14910}\mainline{5.Rc2} \dtmw{24},
                \restoregame{gba14910}\mainline{5.Rd2} \dtmw{21},
                \restoregame{gba14910}\mainline{5.Bxb4} \dtmw{31},
                \restoregame{gba14910}\mainline{5.Bd2} \dtmw{9},
                \restoregame{gba14910}\mainline{5.Bd4} \dtmw{12},
                \restoregame{gba14910}\mainline{5.Bxe5+} \dtmw{18},
                \restoregame{gba14910}\mainline{5.Qc2} \dtmw{20},
                \restoregame{gba14910}\mainline{5.Qd2} \dtmw{17},
                \restoregame{gba14910}\mainline{5.Qf3} \dtmw{22},
                \restoregame{gba14910}\mainline{5.Kf3} \dtmw{21}\color{black})
          \begin{itemize}
             \item  \varid{1}{2.5.1} \restoregame{gba14910} 
            \mainline{5.fxe5+ Bxe5 6.Rxb4 Nxb4 7.Bxb4 d4} \draw this 
            endgame is completly draw since a couple of pawns will be
            excanged and the remaing ones are mutually blocked. 

            \item  \varid{1}{2.5.2} \restoregame{gba14910} 
            \mainline{5.Rxb4 Bxb4 6.fxe5+ Nxe5 7.Bxb4 Qb6} \draw. 
          \end{itemize}

         \item \varid{1}{2.6} \restoregame{gba149}\mainline{4.Qf3 d4}
              \storegame{gba14913}(\color{red}
              \restoregame{gba14913}\mainline{5.exd4} \dtmw{20},
              \restoregame{gba14913}\mainline{5.e4} \dtmw{10},
              \restoregame{gba14913}\mainline{5.Rb3} \dtmw{10},
              \restoregame{gba14913}\mainline{5.Rxb4} \dtmw{20},
              \restoregame{gba14913}\mainline{5.Nxb4} \dtmw{10},
              \restoregame{gba14913}\mainline{5.Nxd4} \dtmw{10},
              \restoregame{gba14913}\mainline{5.Bxb4} \dtmw{19},
              \restoregame{gba14913}\mainline{5.Bd2} \dtmw{22},
              \restoregame{gba14913}\mainline{5.Bxd4} \dtmw{15},
              \restoregame{gba14913}\mainline{5.Qe2} \dtmw{10},
              \restoregame{gba14913}\mainline{5.Qe4} \dtmw{8},
             \restoregame{gba14913}\mainline{5.Qd5} \dtmw{9},
             \restoregame{gba14913}\mainline{5.Qxc6} \dtmw{11}, 
              \restoregame{gba14913}\mainline{5.Ke2} \dtmw{9}\color{black})
   
              \restoregame{gba14913}\mainline{5.fxe5+ Nxe5 6.Bxd4 b3} 
             \wloss \justif \mainline{7.Ke2 bxc2=Q+ 8.Rxc2 Qb3} White
             can hold the balance due to the pin on the Knight and of
             the threat e4 which forces Black to move back his Queen
             to e6.
    \end{itemize}

     \item \varid{1}{3} \restoregame{gba1}\mainline{2.Nxb4 Nxb4}
         \storegame{gba17}
         (\color{red}\restoregame{gba17}  \mainline{3.c4 bxc4} \dtmw{12}, 
          \restoregame{gba17}  \mainline{3.e4 Nc6} \dtmw{33}, 
          \restoregame{gba17}  \mainline{3.f4 Nc6} \dtmw{33}, 
          \restoregame{gba17}  \mainline{3.Rb3 d4} \dtmw{18}, 
          \restoregame{gba17}  \mainline{3.Rxb4 Bxb4} \dtmw{25},
          \restoregame{gba17}  \mainline{3.Rc2 Nxc2} \dtmw{11} \color{black}) 
       
         \begin{itemize}   
           \item \varid{1}{3.1}
          \restoregame{gba17}  \mainline{3.cxb4 d4}
           \storegame{gba171} 
           (\color{red}\restoregame{gba171}\mainline{4.Rb3 Qxb3} \dtmw{12},
            \restoregame{gba171}\mainline{4.Rc2 Qb3} \dtmw{16}, 
            \restoregame{gba171}\mainline{4.Bc3 dxc3} \dtmw{11}\color{black}) 
        \begin{itemize}
          \item \varid{1}{3.1.1} \restoregame{gba171}\mainline{4.exd4
            exd4} \storegame{gba1711}
          (\color{red}\restoregame{gba1711}\mainline{5.Rb3} \dtmw{9},
          \restoregame{gba1711}\mainline{5.Rc2} \dtmw{13},
          \restoregame{gba1711}\mainline{5.Bc3} \dtmw{11},
          \restoregame{gba1711}\mainline{5.Be3} \dtmw{29}, 
          \restoregame{gba1711}\mainline{5.Bf4} \dtmw{15},
          \restoregame{gba1711}\mainline{5.Qe3} \dtmw{8},
          \restoregame{gba1711}\mainline{5.Qe4} \dtmw{10},
          \restoregame{gba1711}\mainline{5.Qe5} \dtmw{7} \color{black})
          \begin{itemize}
          \item \varid{1}{3.1.1.1}\restoregame{gba1711}\mainline{5.f4
              Qxe2+ 6.Kxe2 Rc6} \draw the position is totally blocked
            on dark squarres and White can only play his King or his
            Rook on b2 b3.

          \item \varid{1}{3.1.1.2} \restoregame{gba1711}
            \mainline{5.Qxe6+ Kxe6} \draw for the same reasons as line
            \varid{1}{3.1.1}. 
          \end{itemize}
 
          \item \varid{1}{3.1.2} \restoregame{gba171}\mainline{4.e4
              f4} \storegame{gba17112} 
            (\color{red}\restoregame{gba17112}\mainline{5.Rb3} \dtmw{6}, 
             \restoregame{gba17112}\mainline{5.Rc2} \dtmw{16}, 
             \restoregame{gba17112}\mainline{5.Bc3} \dtmw{8}, 
             \restoregame{gba17112}\mainline{5.Be3} \dtmw{10}, 
             \restoregame{gba17112}\mainline{5.Qe3} \dtmw{9}\color{black})
 
            the only move leads to a type of drawn position already 
            seen in line \varid{1}{1.2} of the White oracle
             \restoregame{gba17112}\mainline{5.Bxf4 exf4 6.Qc4 Qe5} 
              \draw White can just move his Rook on b2-b3 or his
              King other over moves are loosing (he cannot give up the
              control of the c file). 
              
          \item \varid{1}{3.1.3} \restoregame{gba171}\mainline{4.f4 
             exf4} \storegame{gba1713} 
             (\color{red}\restoregame{gba1713}\mainline{5.e4} \dtmw{29}, 
              \restoregame{gba1713}\mainline{5.Rb3} \dtmw{12}, 
              \restoregame{gba1713}\mainline{5.Rc2} \dtmw{17}, 
              \restoregame{gba1713}\mainline{5.Bc3} \dtmw{2}, 
              \restoregame{gba1713}\mainline{5.Qf3} \dtmw{19},
              \restoregame{gba1713}\mainline{5.Kf3} \dtmw{12}\color{black}) 
             \begin{itemize}
               \item \varid{1}{3.1.3.1} \restoregame{gba1713}
                 \mainline{5.exd4 Qxe2+ 6.Kxe2 Ke6} \draw Black King
                 will seat on d5 and White cannot get through.

               \item \varid{1}{3.1.3.2} \restoregame{gba1713}
                 \mainline{5.exf4 Qxe2+ 6.Kxe2 Ke6} \draw same as line
                 above, the Black King seats on d5 and Black may just
                 move his Rook between b6 c6.
               \end{itemize}

       \item   \varid{1}{3.2} \restoregame{gba17}  \mainline{3.d4 e4}
              \storegame{gba14} 
              (\color{red}\restoregame{gba14}\mainline{4.c4} \dtmw{17},
               \restoregame{gba14}\mainline{4.f4} \dtmw{10},
               \restoregame{gba14}\mainline{4.Rxb4}  \dtmw{9}, 
               \restoregame{gba14}\mainline{4.Rb3}  \dtmw{9},
               \restoregame{gba14}\mainline{4.Qxb5} \dtmw{8}, 
               \restoregame{gba14}\mainline{4.Rc2} \dtmw{8}, 
               \restoregame{gba14}\mainline{4.Qc4} \dtmw{8}\color{black}) 

              \restoregame{gba14} \mainline{4.cxb4 exf3 } \storegame{gba141}
              (\color{red}\restoregame{gba141}\mainline{5.Qd3} \dtmw{33},
               \restoregame{gba141}\mainline{5.Rc2} \dtmw{9},
               \restoregame{gba141}\mainline{5.Qxb5}  \dtmw{10}, 
               \restoregame{gba141}\mainline{5.Rb3}  \dtmw{9},
               \restoregame{gba141}\mainline{5.Bc3} \dtmw{8}, 
               \restoregame{gba141}\mainline{5.e4} \dtmw{7}, 
               \restoregame{gba141}\mainline{5.Qc4} \dtmw{7}\color{black}) 
              \begin{itemize}
                 \item   \varid{1}{3.2.1}  
                   \restoregame{gba141}\mainline{5.Qxf3 Rc6}  \storegame{gba1411} 
                 (\color{red}\restoregame{gba1411}\mainline{6.Rb3} \dtmw{31},
                  \restoregame{gba1411}\mainline{6.e4} \dtmw{14},
                  \restoregame{gba1411}\mainline{6.Bc3} \dtmw{12},  
                  \restoregame{gba1411}\mainline{6.Qf4} \dtmw{9},
                  \restoregame{gba1411}\mainline{6.Qxd5} \dtmw{8},
		  \restoregame{gba1411}\mainline{6.Rc2} \dtmw{7}\color{black})
                 
                  \subitem - \varid{1}{3.2.1.1}\restoregame{gba1411}\mainline{6.Qe2 Rc4}   
                  \storegame{gba14111} 
                 (\color{red}\restoregame{gba14111}\mainline{7.Rb3} \dtmw{29},
                  \restoregame{gba14111}\mainline{7.Kf3} \dtmw{27},
                   \restoregame{gba14111}\mainline{7.Qxc4} \dtmw{17}\color{black})
                 
                  \subsubitem * \varid{1}{3.2.1.1.1}\restoregame{gba14111}\mainline{7.Qf3 Qe4}    
                      \storegame{gba141111} 
                      (\color{red}\restoregame{gba141111}\mainline{8.Ke2} \dtmw{32},
                      \restoregame{gba141111}\mainline{8.Qe2} \dtmw{25},
                      \restoregame{gba141111}\mainline{8.Rb3}
                      \dtmw{21}\color{black})

		      \restoregame{gba141111}\mainline{8.Qxe4 fxe4} \draw
                      White cannot get through since his Bishope is
                      limited by his pawns. If the White Rook moves to
                      the third raw then \bmove{Rc2} limits the White choice
                      to \wmove{Rc3} after the Rook exchange the
                      position is an easy draw.   
                      
                   \subsubitem *  \varid{1}{3.2.1.1.2}
                      \restoregame{gba14111}\mainline{7.Qd3 Qe4} \draw
                      if White takes on e4 we have the same position
                      as variation \varid{1}{3.2.1.1.1} otherwise
                      Black just moves his king on e6-f6.

                  \subitem - \varid{1}{3.2.1.2}  \restoregame{gba1411}
                  \mainline{6.Ke2 Rc4} \draw similarly to lines 
                  \varid{1}{3.2.1.1.1} and \varid{1}{3.2.1.1.2} Black
                  will play \bmove{Qe4} and block the position. 

              \item   \varid{1}{3.2.2}  \restoregame{gba141}
                 \mainline{5.Kxf3 Qe4 6.Kf2 Ke6} \storegame{gba141a}
                 (\color{red}\restoregame{gba141a} \mainline{7.Qc4} \dtmw{7}, 
                 \restoregame{gba141a} \mainline{7.Qxb5} \dtmw{8},
                 \restoregame{gba141a} \mainline{7.Qd3} \dtmw{8}, 
                 \restoregame{gba141a} \mainline{7.Rc2} \dtmw{12},
                 \restoregame{gba141a} \mainline{7.Qf3 Rc6} \draw see
                 line \varid{1}{3.2.1.1.2}\color{black})

                  \restoregame{gba141a} \mainline{7.Rb3 Rc6} 
                   (\color{red}\mainline{8.Qxb5}
                   \dtmw{34}\color{black}) \draw
                  if White does not take the b5 pawn the position is 
                  similar to line \varid{1}{3.2.1.1.1}.

	  \end{itemize}	       
 \end{itemize}
\end{itemize}
\end{itemize}

\restoregame{gba1}\mainline{2.cxb4 d4} \storegame{gba2}
(\color{red}\restoregame{gba2}\mainline{3.Bc3 dxc3} \dtmw{8},  
 \restoregame{gba2}\mainline{3.Rb3 Qxb3}\dtmw{11}\color{black})

\begin{itemize}
   \item \varid{2}{1} \restoregame{gba2}\mainline{3.exd4 exd4}
         \storegame{gba21} 
         (\color{red}\restoregame{gba21}\mainline{4.Rb3} \dtmw{10},
          \restoregame{gba21}\mainline{4.Ne3} \dtmw{29}, 
          \restoregame{gba21}\mainline{4.Bc3} \dtmw{15},
          \restoregame{gba21}\mainline{4.Be3} \dtmw{19}, 
          \restoregame{gba21}\mainline{4.Bf4} \dtmw{11}\color{black})
         \begin{itemize}
           \item \varid{2}{1.1} \restoregame{gba21}\mainline{4.f4
               Qxe2+ 5.Kxe2 f4} \draw since only Kings can move without
             losing a piece and leading to a lost position
             (\wmove{Rb3} is possible but changes nothing to the
             evaluation of the position).

           \item \varid{2}{1.2} \restoregame{gba21}\mainline{4.Qxe6+
               Kxe6 5.f4}  \draw similar as line \varid{2}{1.1}. 
               \item \varid{2}{1.3} \restoregame{gba21}
                 \mainline{4.Nxd4 Nxd4} \storegame{gba213}  
                (\color{red}\restoregame{gba213}\mainline{5.Qe3}
                 \dtmw{22}, \restoregame{gba213}\mainline{5.Qxe6}
                 \dtmw{16}, \restoregame{gba213}\mainline{5.Be3}
                 \dtmw{10}\color{black})

                 \restoregame{gba213}\mainline{5.Bc3}
           \end{itemize}

   \item \varid{2}{2} \restoregame{gba2}\mainline{3.f4 exf4}
     \storegame{gba22} 
     (\color{red}\restoregame{gba21}\mainline{4.e4} \dtmw{24}, 
     \restoregame{gba21}\mainline{4.Rb3} \dtmw{14},
     \restoregame{gba21}\mainline{4.Bc3} \dtmw{11},
     \restoregame{gba21}\mainline{4.Qf3} \dtmw{23},
     \restoregame{gba21}\mainline{4.Kf3} \dtmw{15}\color{black}) 
     \begin{itemize}
       \item \varid{2}{2.1} \restoregame{gba22}\mainline{4.exd4 Qxe2+
               5.Kxe2 Ke6} \draw the Black King will move to d5-e6. 

       \item \varid{2}{2.2} \restoregame{gba22}\mainline{4.exf4 Qxe2+
              5.Kxe2} \draw see variation \varid{2}{1.2}. 

          \item \varid{2}{2.3} \restoregame{gba22}\mainline{4.Nxd4
              Nxd4 5.exd4} (\variation{5.Bc3 fxe3+ 6.Qxe3 Qxe3+ 7.Kxe3
              Be5} followed by exchanges to a completly drawn endgame
            does not change the assesment of the position)
            \mainline{5... Qxe2+ 6.Kxe2 Ke6} \draw the Black King will
            seat on d5.
       \end{itemize}

   \item \varid{2}{4} \restoregame{gba2}\mainline{3.Nxd4 Nxd4}
          \storegame{gba24} 

   \begin{itemize}
     \item \varid{2}{4.1} \restoregame{gba24}\mainline{4.exd4 exd4}
            (\color{red}
            \storegame{gba241} \mainline{5.Rb3} \dtmw{9},
          \restoregame{gba241}\mainline{5.Rc2} \dtmw{13},
          \restoregame{gba241}\mainline{5.Bc3} \dtmw{11},
          \restoregame{gba241}\mainline{5.Be3} \dtmw{19},
          \restoregame{gba241}\mainline{5.Bf4} \dtmw{15},
          \restoregame{gba241}\mainline{5.Qe3} \dtmw{8},
          \restoregame{gba241}\mainline{5.Qe4} \dtmw{10},
          \restoregame{gba241}\mainline{5.Qe5} \dtmw{7}\color{black})
     \begin{itemize}
       \item \varid{2}{4.1.1} \restoregame{gba241}\mainline{5.f4
           Qxe2+ 6.Kxe2} \draw see variation \varid{1}{3.1.1.1}
 
       \item \varid{2}{4.1.2} \restoregame{gba241}\mainline{5.Qxe6+
           Kxe6} \draw see variation \varid{1}{3.1.1.2}
     \end{itemize}
   \end{itemize}
\end{itemize}

\restoregame{gba2}\mainline{3.e4 f4} \storegame{gba3}
(\color{red}\restoregame{gba3}\mainline{4.Rb3  Qxb3} 
          \dtmw{7},
 \restoregame{gba3}\mainline{4.Nxd4 Nxd4}
          \dtmw{16},
\restoregame{gba3}\mainline{4.Ne3 fxe3} 
          \dtmw{18},
\restoregame{gba3}\mainline{4.Bc3 dxc3} 
         \dtmw{9},
\restoregame{gba3}\mainline{4.Be3} 
         \dtmw{18},
\restoregame{gba3}\mainline{4.Qe3} 
         \dtmw{12}\color{black})

\restoregame{gba3}\mainline{4.Bxf4 exf4} \storegame{gba4}
(\color{red}\restoregame{gba4}\mainline{5.e5} 
          \dtmw{16},
\restoregame{gba4}\mainline{5.Rb3} 
          \dtmw{7},
 \restoregame{gba4}\mainline{5.Nxd4} 
          \dtmw{9},
\restoregame{gba4}\mainline{5.Ne3} 
          \dtmw{7},
 \restoregame{gba4}\mainline{5.Qe3} 
          \dtmw{7}\color{black})

\restoregame{gba4}\mainline{5.Qd2 Be5} \storegame{gba6} \draw
since the only non losing moves for White are limited to the King and
Queen moves over the d2, e2 and f2 squares. 

   \subsection{White moves \wmove{c4}}
\gardnerstart
\mainline{1.c4 bxc4} The pin on the b file leads to forced mate \dtmw{27}.

   \subsection{White moves \wmove{d4}}
\gardnerstart
\mainline{1.d4 e4}\storegame{gbc1}
(\color{red}\restoregame{gbc1}\mainline{2.Nb4} \dtmw{21},
 \restoregame{gbc1}\mainline{2.Qxb5} \dtmw{12},
 \restoregame{gbc1}\mainline{2.Qc4} \dtmw{9},
 \restoregame{gbc1}\mainline{2.Qd3} \dtmw{10},
 \restoregame{gbc1}\mainline{2.fxe4} \dtmw{13}\color{black})
\begin{itemize}
  \item \varid{1}{1} 
        \restoregame{gbc1}\mainline{2.b4 c4} \storegame{gbc11}
        (\color{red}\restoregame{gbc11}\mainline{3.fxe4} \dtmw{9}, 
         \restoregame{gbc11}\mainline{3.Rb3} \dtmw{8},
         \restoregame{gbc11}\mainline{3.Qxc4}\dtmw{16},
         \restoregame{gbc11}\mainline{3.Qd3} \dtmw{9}\color{black})
        \restoregame{gbc11} \mainline{3.f4 Bxb4}
        \storegame{gbc11a}
        (\color{red}\restoregame{gbc11a}\mainline{4.Rb3} \dtmw{8}, 
        \restoregame{gbc11a}\mainline{4.Rxb4 Nxb4} \dtmw{21},
        \restoregame{gbc11a}\mainline{4.Qxc4} \dtmw{15},
        \restoregame{gbc11a}\mainline{4.Qd3} \dtmw{10},
        \restoregame{gbc11a}\mainline{4.Qf3} \dtmw{11}\color{black})
  \begin{itemize}
    \item \varid{1}{1.1}\restoregame{gbc11a}\mainline{4.cxb4 Qd6}
      \draw despite his extra piece White cannot win since he is
      blocked by his own pawns on dark squares.
    \item \varid{1}{1.2}\restoregame{gbc11a}\mainline{4.Nxb4 Nxb4} \draw
          \justif \storegame{gbc11a4}
          \mainline{5.Rxb4 Qd6} and White may only move his Rook, on 
          \restoregame{gbc11a4}\mainline{5.bxc4 Qd6} is similar to 
          \varid{1}{1.1}. 
  \end{itemize}

  \item \varid{1}{2} \restoregame{gbc1} 
        \mainline{2.c4 bxc4} \storegame{gbc12}
        (\color{red}\restoregame{gbc12}\mainline{3.b4} \dtmw{15},
         \restoregame{gbc12} \mainline{3.bxc4} \dtmw{19},
         \restoregame{gbc12} \mainline{3.Nb4} \dtmw{14},
         \restoregame{gbc12} \mainline{3.Bb4} \dtmw{10},
         \restoregame{gbc12} \mainline{3.fxe4} \dtmw{10},
         \restoregame{gbc12} \mainline{3.f4} \dtmw{17},
         \restoregame{gbc12} \mainline{3.Qd3} \dtmw{8},
         \restoregame{gbc12} \mainline{3.Qxc4} \dtmw{12}\color{black})
        \begin{itemize}
          \item \varid{1}{2.1} \restoregame{gbc12} 
                 \mainline{3.dxc5 Bxc5} \storegame{gbc121a}
                 (\color{red}\restoregame{gbc121a}\mainline{4.b4} \dtmw{20},
                  \restoregame{gbc121a} \mainline{4.Nb4} \dtmw{12},
                  \restoregame{gbc121a} \mainline{4.Nd4} \dtmw{15},
                  \restoregame{gbc121a} \mainline{4.Bb4} \dtmw{14},
                  \restoregame{gbc121a} \mainline{4.Qd3} \dtmw{8},
                  \restoregame{gbc121a} \mainline{4.Qxc4} \dtmw{11},
                  \restoregame{gbc121a} \mainline{4.f4} \dtmw{12}\color{black})

                  \restoregame{gbc121a} \mainline{4.Bc3 Qe5} 
                  \storegame{gbc121b}
                  (\color{red}\restoregame{gbc121b}\mainline{5.Bxe5+} \dtmw{19},
                  \restoregame{gbc121b} \mainline{5.bxc4} \dtmw{16}, 
                  \restoregame{gbc121b} \mainline{5.fxe4} \dtmw{13}, 
                  \restoregame{gbc121b} \mainline{5.b4} \dtmw{9},
                  \restoregame{gbc121b} \mainline{5.Nd4} \dtmw{12},
                  \restoregame{gbc121b} \mainline{5.Nb4} \dtmw{12},
                  \restoregame{gbc121b} \mainline{5.f4} \dtmw{7},
                  \restoregame{gbc121b} \mainline{5.Bd4} \dtmw{9},
                  \restoregame{gbc121b} \mainline{5.Qxc4} \dtmw{9},
                  \restoregame{gbc121b} \mainline{5.Qc3} \dtmw{7}\color{black})

                  \restoregame{gbc121b} \mainline{5.Qd2 f4} 
                  \storegame{gbc121c}   
                  (\color{red}\restoregame{gbc121c}\mainline{6.Ke2} \dtmw{12},
                   \restoregame{gbc121c} \mainline{6.fxe4} \dtmw{13},
                   \restoregame{gbc121c} \mainline{6.b4} \dtmw{17},
                   \restoregame{gbc121c} \mainline{6.Nd4} \dtmw{10},
                   \restoregame{gbc121c} \mainline{6.Bd4} \dtmw{8},
                   \restoregame{gbc121c} \mainline{6.bxc4} \dtmw{8},
                   \restoregame{gbc121c} \mainline{6.Nb4} \dtmw{8},
                   \restoregame{gbc121c} \mainline{6.Qd4} \dtmw{7},
                   \restoregame{gbc121c} \mainline{6.Bb4} \dtmw{7},
                   \restoregame{gbc121c} \mainline{6.Qe2} \dtmw{6},
                   \restoregame{gbc121c} \mainline{6.Qxd5} \dtmw{6},
                   \restoregame{gbc121c} \mainline{6.Qd3} \dtmw{5}\color{black})
 
                   \restoregame{gbc121c} \mainline{6.Bxe5 Nxe5}
                   \storegame{gbc121d}
                   (\color{red}\restoregame{gbc121d}\mainline{7.Nd4} \dtmw{16}, 
                    \restoregame{gbc121d} \mainline{7.Qc3}\dtmw{12},
                    \restoregame{gbc121d} \mainline{7.fxe4} \dtmw{20},
                    \restoregame{gbc121d} \mainline{7.Qd4} \dtmw{11},
                    \restoregame{gbc121d} \mainline{7.Ke2} \dtmw{11},
                    \restoregame{gbc121d} \mainline{7.Nb4} \dtmw{10},
                    \restoregame{gbc121d} \mainline{7.Qxd5} \dtmw{8},
                    \restoregame{gbc121d} \mainline{7.Qb4} \dtmw{6},
                    \restoregame{gbc121d} \mainline{7.bxc4} \dtmw{5},
                    \restoregame{gbc121d} \mainline{7.Qe2} \dtmw{5},
                    \restoregame{gbc121d} \mainline{7.Qd3} \dtmw{4}\color{black})

                  \restoregame{gbc121d} \mainline{7.b4 fxe3+}
                  \storegame{gbc121e}
                   (\color{red}\restoregame{gbc121e}\mainline{8.Qxe3} \dtmw{14}, 
                    \restoregame{gbc121e} \mainline{8.Ke2} \dtmw{1}\color{black})
                 
                  \restoregame{gbc121e} \mainline{8.Nxe3 Nd3+}
                  \storegame{gbc121f}
                  (\color{red}\restoregame{gbc121f}\mainline{9.Qxd3} \dtmw{6}\color{black})
   
                  \restoregame{gbc121f} \mainline{9.Ke2 Nf4+ 10.Kf2
                    Nf4+} \draw draw by repetition.
 
          \item \varid{1}{2.2} \restoregame{gbc12}
                \mainline{3.Bc3 Rxb3} \storegame{gbc122a}
 
                \restoregame{gbc122a} \mainline{4.Rxb3 cxb3}
                \storegame{gbc122b}
                (\color{red}\restoregame{gbc122b}\mainline{5.fxe4} \dtmw{9},
                \restoregame{gbc122b}\mainline{5.f4} \dtmw{15},
                \restoregame{gbc122b}\mainline{5.Nb4} \dtmw{11},
                \restoregame{gbc122b}\mainline{5.Bb4} \dtmw{8},
                \restoregame{gbc122b}\mainline{5.Bb2} \dtmw{10},
                \restoregame{gbc122b}\mainline{5.Bd2} \dtmw{6},
                \restoregame{gbc122b}\mainline{5.Qb5} \dtmw{6},
                \restoregame{gbc122b}\mainline{5.Qc4} \dtmw{6},
                \restoregame{gbc122b}\mainline{5.Qd3} \dtmw{5},
                \restoregame{gbc122b}\mainline{5.Qd2} \dtmw{15}\color{black})

                \restoregame{gbc122b} \mainline{5.dxc5+ Be5}
                \storegame{gbc122c}
                (\color{red}\restoregame{gbc122c}\mainline{6.fxe4} \dtmw{8},
                \restoregame{gbc122c}\mainline{6.f4} \dtmw{6},
                \restoregame{gbc122c}\mainline{6.Nb4} \dtmw{7},
                \restoregame{gbc122c}\mainline{6.Bb2} \dtmw{8},
                \restoregame{gbc122c}\mainline{6.Bb4} \dtmw{9},
                \restoregame{gbc122c}\mainline{6.Bd2} \dtmw{7},
                \restoregame{gbc122c}\mainline{6.Bd4} \dtmw{9},
                \restoregame{gbc122c}\mainline{6.Qd2} \dtmw{16},
                \restoregame{gbc122c}\mainline{6.Qd3} \dtmw{4},
                \restoregame{gbc122c}\mainline{6.Qc4} \dtmw{6},
                \restoregame{gbc122c}\mainline{6.Qb5} \dtmw{6},
                \restoregame{gbc122c}\mainline{6.Bxe5+} \dtmw{13}\color{black})

                \restoregame{gbc122c} \mainline{6.Nd4 Bxd4}
                \storegame{gbc122d}
                (\color{red}\restoregame{gbc122d}\mainline{7.fxe4} \dtmw{7},
                \restoregame{gbc122d} \mainline{7.f4} \dtmw{6},
                \restoregame{gbc122d} \mainline{7.Bb2} \dtmw{11},
                \restoregame{gbc122d} \mainline{7.Bb4} \dtmw{5},
                \restoregame{gbc122d} \mainline{7.Bd2} \dtmw{6},
                \restoregame{gbc122d} \mainline{7.Qd2} \dtmw{11},
                \restoregame{gbc122d} \mainline{7.Qd3} \dtmw{3},
                \restoregame{gbc122d} \mainline{7.Qc4} \dtmw{8},
                \restoregame{gbc122d} \mainline{7.Qb5} \dtmw{10},
                \restoregame{gbc122d} \mainline{7.Qb2} \dtmw{11},
                \restoregame{gbc122d} \mainline{7.Qc2} \dtmw{2}\color{black})

                   -  \varid{1}{2.2.1}\restoregame{gbc122d} 
                          \mainline{7.Bxd4+ Nxd4}
                          \storegame{gbc1221}
                          (\color{red}\restoregame{gbc1221}\mainline{8.Qd2}\dtmw{11},
                          \restoregame{gbc1221}\mainline{8.fxe4}\dtmw{6},
                          \restoregame{gbc1221}\mainline{8.c6=Q}\dtmw{5}
                          (other promotions as well),
                          \restoregame{gbc1221}\mainline{8.f4}\dtmw{3},
                          \restoregame{gbc1221}\mainline{8.Qc2
                            bxc2=Q+} checkmate \color{black})

                          \restoregame{gbc1221}\mainline{8.Qb2 f4}
                          \storegame{gbc1221a}
                          (\color{red}\restoregame{gbc1221a}\mainline{9.Qxd4+}\dtmw{17},
                          \restoregame{gbc1221a}\mainline{9.c6=Q}\dtmw{11}
                          (other promotions as well),
                          \restoregame{gbc1221a}\mainline{9.fxe4}\dtmw{7},
                          \restoregame{gbc1221a}\mainline{9.Qd2}\dtmw{6},
                          \restoregame{gbc1221a}\mainline{9.Qc2
                            bxc2=Q+} checkmate \color{black})

                          \restoregame{gbc1221a}\mainline{9.exd4 exf3}
                          \storegame{gbc1221b}
                          (\color{red}\restoregame{gbc1221b}\mainline{10.Qd2}\dtmw{3},
                          \restoregame{gbc1221b}\mainline{10.c6=B Qe3}
                          checkmate,
                          \restoregame{gbc1221b}\mainline{10.c6=N Qe3}
                          checkmate,
                          \restoregame{gbc1221b}\mainline{10.Qc2 Qe3}
                          checkmate,
                          \restoregame{gbc1221b}\mainline{10.Qe2 fxe2=Q+}
                          checkmate,
                          \restoregame{gbc1221b}\mainline{10.Qxb3 Qe2}
                          checkmate, 
                          \restoregame{gbc1221b}\mainline{10.Qc3 Qe2}
                          checkmate \color{black})

                          \restoregame{gbc1221b}\mainline{10.c6=Q
                            Qxc6}, promotion to Rook is handled similarly,
                          \storegame{gbc1221c}
                          (\color{red}\restoregame{gbc1221c}\mainline{11.Kxf3}\dtmw{7},
                          \restoregame{gbc1221c}\mainline{11.Qc3}\dtmw{4},
                          \restoregame{gbc1221c}\mainline{11.Qd2}\dtmw{4},
                          \restoregame{gbc1221c}\mainline{11.Qc2}\dtmw{3},
                          \restoregame{gbc1221c}\mainline{11.Qe2}\dtmw{3}\color{black})

                          \restoregame{gbc1221c}\mainline{11.Qxb3 Qe6}
                          \draw Black will play \bmove{Qe2+} and after
                          Queen exchange the pawn endgame is draw.

                   - \varid{1}{2.2.2}\restoregame{gbc122d} 
                         \mainline{7.exd4 e3+ 8.Qxe3 Qxe3+ 9.Kxe3}
                           \storegame{gbc1222}  \draw the Black King
                           just moves to e6-f6 and White King cannot 
                           break through. If the White Biwhop goes to
                           e5 either Black can play f4 and get room
                           for his King or it means that White played
                           f4 hence after \bmove{Nb4} the Knight
                           cannot be taken without stalemating the
                           Black King. 

  \item \varid{1}{3} \restoregame{gbc1}\mainline{2.dxc5 Bxc5} 
         \storegame{gbc13}
         (\color{red}\restoregame{gbc13}\mainline{3.c4 b4}\dtmw{23},
          \restoregame{gbc13}\mainline{3.fxe4} \dtmw{14}, 
          \restoregame{gbc13}\mainline{3.f4 Qd6}\dtmw{14}, 
          \restoregame{gbc13}\mainline{3.Nb4 f4}\dtmw{16},
          \restoregame{gbc13}\mainline{3.Qxb5} \dtmw{11},
          \restoregame{gbc13}\mainline{3.Qc4} \dtmw{9},
          \restoregame{gbc13}\mainline{3.Qd3} \dtmw{9},
          \restoregame{gbc13}\mainline{3.b4} \dtmw{16}\color{black})
 
         \restoregame{gbc13}\mainline{3.Nd4 Nxd4}
           \storegame{gbc131}
          (\color{red}\restoregame{gbc131}\mainline{4.b4} \dtmw{12},  
           \restoregame{gbc131}\mainline{4.c4} \dtmw{9}, 
                  \restoregame{gbc131}\mainline{4.fxe4}
               \dtmw{10},  \restoregame{gbc131}\mainline{4.f4}
                \dtmw{9}, \restoregame{gbc131}\mainline{4.Rc2}
                \dtmw{9},  \restoregame{gbc131}\mainline{4.Qxb5}
                 \dtmw{8}, \restoregame{gbc131}\mainline{4.Qc4}
                \dtmw{8}, \restoregame{gbc131}\mainline{4.Qd3}
                 \dtmw{7}\color{black})
         \begin{itemize}
           \item \varid{1}{3.1} \restoregame{gbc131}
                  \mainline{4.cxd4 exf3} \storegame{gbc1312}
                  (\color{red}\restoregame{gbc1312}\mainline{5.b4} \dtmw{12},
                  \restoregame{gbc1312}\mainline{5.dxc5} \dtmw{14},
                  \restoregame{gbc1312}\mainline{5.e4} \dtmw{5},
                  \restoregame{gbc1312}\mainline{5.Rc2} \dtmw{10},
                  \restoregame{gbc1312}\mainline{5.Bc3} \dtmw{10},
                  \restoregame{gbc1312}\mainline{5.Bb4} \dtmw{8},
                  \restoregame{gbc1312}\mainline{5.Qc4} \dtmw{12},
                  \restoregame{gbc1312}\mainline{5.Qxb5} \dtmw{16}\color{black})
                  \begin{itemize}
                    \item \varid{1}{3.1.1}
                      \restoregame{gbc1312}\mainline{5.Qd3 Bd6}  \storegame{gbc1312a}
                      \draw
                      on any reasonable move 
                      (\color{red}\restoregame{gbc1312a}\mainline{6.Qe4} \dtmw{6}, 
                       \restoregame{gbc1312a}\mainline{6.Qc2} \dtmw{32}, 
                       \restoregame{gbc1312a}\mainline{6.b4} \dtmw{34}, 
                       \restoregame{gbc1312a}\mainline{6.Bc3} \dtmw{24}, 
                       \restoregame{gbc1312a}\mainline{6.Bb4} \dtmw{15}, 
                       \restoregame{gbc1312a}\mainline{6.Qc3} \dtmw{24}, 
                       \restoregame{gbc1312a}\mainline{6.e4} \dtmw{12}, 
                       \restoregame{gbc1312a}\mainline{6.Qxb5} \dtmw{8}, 
                       \restoregame{gbc1312a}\mainline{6.Qe2} \dtmw{8}, 
                       \restoregame{gbc1312a}\mainline{6.Qc4} \dtmw{7}, 
                       \restoregame{gbc1312a}\mainline{6.Qxf5+} \dtmw{6}\color{black}) 
                      Black plays \bmove{Qe4} and
                      locks the position as in variation 
                      \varid{1}{3.1.3.1}.

                    \item \varid{1}{3.1.2}
                      \restoregame{gbc1312}\mainline{5.Qxf3 Bd6} 
                      \storegame{gbc1312b}\draw 
                      on any reasonable move 
                     (\color{red}\restoregame{gbc1312b}\mainline{6.Qxf5+} \dtmw{7},
                      \restoregame{gbc1312b}\mainline{6.Qf4} \dtmw{8},
                      \restoregame{gbc1312b}\mainline{6.Qxd5} \dtmw{9},
                      \restoregame{gbc1312b}\mainline{6.Qe4} \dtmw{4}
                      \restoregame{gbc1312b}\mainline{6.e4} \dtmw{37},
                      \restoregame{gbc1312b}\mainline{6.Bb4} \dtmw{22}\color{black})
                      Black plays \bmove{Qe4} and
                      locks the position as in variation 
                      \varid{1}{3.2.3.1}.

                    \item \varid{1}{3.1.3}
                      \restoregame{gbc1312}\mainline{5.Kxf3 Qe4+ 6.Kf2
                      Bd6} \storegame{gbc13123}
                     (\color{red}\restoregame{gbc13123}\mainline{7.Rc2} \dtmw{10},
                     \restoregame{gbc13123}\mainline{7.Bb4} \dtmw{12},
                     \restoregame{gbc13123}\mainline{7.Qd3} \dtmw{7},
                     \restoregame{gbc13123}\mainline{7.Qc4} \dtmw{6},
                     \restoregame{gbc13123}\mainline{7.Qxb5} \dtmw{8}\color{black})

                     \subitem
                     \varid{1}{3.1.3.1}\restoregame{gbc13123}\mainline{7.b4
                     Ke6} \draw Black just moves his King on e6-f6 and
                   the position is blocked on the dark squares 
                   \justif \mainline{8.Qf2 Kf6 9.Qxe3 fxe3}

                     \subitem
                     \varid{1}{3.1.3.2}\restoregame{gbc13123}\mainline{7.Bc3
                     Ke6} \draw see line \varid{1}{3.2.3.1}.

                     \subitem
                     \varid{1}{3.1.3.3}\restoregame{gbc13123}\mainline{7.Qf3
                     Ke6} \draw see line \varid{1}{3.2.3.1}.

                  \end{itemize}

           \item \varid{1}{3.2} \restoregame{gbc131}\mainline{4.exd4 Bd6} 
                 \storegame{gbc1314} 
                 (\color{red}\restoregame{gbc1314}\mainline{5.Rc2 f4} \dtmw{23},
                  \restoregame{gbc1314}\mainline{5.b4 f4} \dtmw{18},
                  \restoregame{gbc1314}\mainline{5.Be3 f4} \dtmw{22},
                  \restoregame{gbc1314}\mainline{5.Qe3 f4} \dtmw{22},
                  \restoregame{gbc1314}\mainline{5.c4 bxc4} \dtmw{23},
                  \restoregame{gbc1314}\mainline{5.Ke3 f4+} \dtmw{23},
                  \restoregame{gbc1314}\mainline{5.Bf4} \dtmw{11},
                  \restoregame{gbc1314}\mainline{5.Qxb5} \dtmw{10},
                  \restoregame{gbc1314}\mainline{5.Qc4} \dtmw{8},
                  \restoregame{gbc1314}\mainline{5.Qd3} \dtmw{4},
                  \restoregame{gbc1314}\mainline{5.Bf4} \dtmw{11}\color{black})
          
                        - \varid{1}{3.2.1} \restoregame{gbc1314} 
                               \mainline{5.fxe4 Qxe4}
                               \draw if White exchanges Queen on e4
                               then with \bmove{fxe4} Black closes the
                               position and with \bmove{Rc6} White
                               cannot do anything. If White does not
                               exchange Queens then Black may just
                               play his King (on  
                               \mainline{6.b4 f4} is \dtmw{28}\color{black}).
                                  
                        - \varid{1}{3.2.2} \restoregame{gbc1314} 
                               \mainline{5.f4 e3+} \draw since Black
                               follows with \bmove{Qe4} and blocks the
                               position.            
         \end{itemize}

  \item \varid{1}{4} \restoregame{gbc1}\mainline{2.f4 c4}
    \storegame{gbc15}
    (\color{red}\restoregame{gbc15}\mainline{3.Qd3} \dtmw{12},
    \restoregame{gbc15}\mainline{3.Qxc4 bxc4} \dtmw{15},    
    \restoregame{gbc15}\mainline{3.Qf3} \dtmw{8}\color{black})    
  \begin{itemize}
    \item \varid{1}{4.1} \restoregame{gbc15}\mainline{3.b4 Bxb4} \draw
          due to the blocked position White cannot achieve anything,
          this type of position has already been treated in 
          line \varid{1}{1} of this oracle for instance.   
    \item \varid{1}{4.2} \restoregame{gbc15}\mainline{3.bxc4 dxc4} 
    \storegame{gbc152}
    (\color{red}\restoregame{gbc152}\mainline{4.d5 Qxd5} \dtmw{22},
     \restoregame{gbc152}\mainline{4.Rxb5 Rxb5} \dtmw{10},
     \restoregame{gbc152}\mainline{4.Rb3} \dtmw{9},
     \restoregame{gbc152}\mainline{4.Qxc4} \dtmw{8},
     \restoregame{gbc152}\mainline{4.Qd3} \dtmw{9},
     \restoregame{gbc152}\mainline{4.Qf3} \dtmw{11},
     \restoregame{gbc152}\mainline{4.Nxb4 Bc5} \dtmw{12} this
         surprising move lead to direct checkmate since White is
         completly blocked and will eventually, due to his lack of
         space, have to gite his Queen within a few moves.\color{black})

      \restoregame{gbc152}
              \mainline{4.Rb4 Nxb4} \storegame{gbc1521}
        (\color{red}\restoregame{gbc1521}\mainline{5.d5} \dtmw{8},
         \restoregame{gbc1521}\mainline{5.Nxb4} \dtmw{23}, 
         \restoregame{gbc1521}\mainline{5.Qxc4} \dtmw{5},
         \restoregame{gbc1521}\mainline{5.Qd3} \dtmw{7},
         \restoregame{gbc1521}\mainline{5.Qf3} \dtmw{6}\color{black})
         \restoregame{gbc1521}\mainline{5.cxb4 Qd5} \draw because the 
         position is totally 
         blocked and Black just moves his King to e6 f6. The only way
         to untangle for White is to sacrifice the Queen on c4 which
         lead to quick checkmate.

   \item \varid{1}{4.3} \restoregame{gbc15}\mainline{3.Nb4 Nxb4}
     \storegame{gbc15a}
     (\color{red}\restoregame{gbc15a} \mainline{4.Rc2} \dtmw{8},
     \restoregame{gbc15a} \mainline{4.Qxc4} \dtmw{9},
     \restoregame{gbc15a} \mainline{4.Qd3} \dtmw{7},
     \restoregame{gbc15a} \mainline{4.bxc4} \dtmw{15},
     \restoregame{gbc15a} \mainline{4.Qf3} \dtmw{7}\color{black})
     \draw The draw is tricky to understand at first sight but becomes
     clear with the following variation 
     \restoregame{gbc15a}\mainline{4.cxb4 Bxf4} \storegame{gbc153}
     (\color{red}\restoregame{gbc153} \mainline{5.Bc3} \dtmw{16},
     \restoregame{gbc153} \mainline{5.Rc2} \dtmw{15},
     \restoregame{gbc153} \mainline{5.Qd3} \dtmw{6},
     \restoregame{gbc153} \mainline{5.Qxc4} \dtmw{9},
     \restoregame{gbc153} \mainline{5.Qf3} \dtmw{7}\color{black}). 
     From here the
     idea is to build a blockade on dark squares. 
       \begin{itemize} 
         \item After \restoregame{gbc153}\mainline{5.exf4 Rc6}
         (in order to be able to take with the Rook in the case of
         \wmove{bxc4}) \draw The blockade has been achieved and Black 
         just moves his Queen on d6 and his King on e6 f6.
       
       \item \restoregame{gbc153}\mainline{5.bxc4 bxc4
           6.exf4 Rb5} \draw another blockade is built on dark squares and
         White cannot break through.
       \end{itemize}
      \end{itemize}
\end{itemize}

   \subsection{White moves \wmove{e4}}
\gardnerstart \mainline{1.e4 f4}\storegame{gbd1} 
(\color{red}\restoregame{gbd1}\mainline{2.d4} \dtmw{25},
\restoregame{gbd1}\mainline{2.Nb4} \dtmw{17},
\restoregame{gbd1}\mainline{2.Nd4} \dtmw{25},
\restoregame{gbd1}\mainline{2.Bxf4 exf4} \dtmw{22},
\restoregame{gbd1}\mainline{2.Qe3 fxe3+} \dtmw{25}\color{black})

\begin{itemize}
\item \varid{1}{1} \restoregame{gbd1}\mainline{2.b4 cxb4}
  \storegame{gbd11}  
  (\color{red}\restoregame{gbd11}\mainline{3.c4} \dtmw{15}, 
  \restoregame{gbd11}\mainline{3.d4} \dtmw{27}, 
  \restoregame{gbd11}\mainline{3.exd5} \dtmw{27}, 
  \restoregame{gbd11}\mainline{3.Rb3} \dtmw{17}, 
  \restoregame{gbd11}\mainline{3.Rxb4} \dtmw{34}, 
  \restoregame{gbd11}\mainline{3.Nxb4} \dtmw{19}, 
  \restoregame{gbd11}\mainline{3.Nd4} \dtmw{14}, 
  \restoregame{gbd11}\mainline{3.Ne3} \dtmw{17}, 
  \restoregame{gbd11}\mainline{3.Be3} \dtmw{14}, 
  \restoregame{gbd11}\mainline{3.Bxf4} \dtmw{16}, 
  \restoregame{gbd11}\mainline{3.Qe3} \dtmw{17}\color{black})  

  \restoregame{gbd11}\mainline{3.cxb4 d4} \draw This position is draw for
  the same reason as position \varid{1}{1.2} of the White oracle (see
  section \ref{sec:oracle_gardner_White}\color{black}). White is in zugzwang and
  must give a piece, the only non losing way to do it is by
  \mainline{4.Bxf4 exf4 5.Qd2}.

  \item \varid{1}{2} \restoregame{gbd1}\mainline{2.c4 bxc4} \storegame{gbd12} 
   (\color{red}\restoregame{gbd12}\mainline{3.b4} \dtmw{22}, 
   \restoregame{gbd12}\mainline{3.dxc4} \dtmw{19}, 
   \restoregame{gbd12}\mainline{3.d4} \dtmw{15}, 
   \restoregame{gbd12}\mainline{3.Nb4} \dtmw{12}, 
   \restoregame{gbd12}\mainline{3.Nd4} \dtmw{15}, 
   \restoregame{gbd12}\mainline{3.Ne3} \dtmw{16}, 
   \restoregame{gbd12}\mainline{3.Bc3} \dtmw{20}, 
   \restoregame{gbd12}\mainline{3.Bb4} \dtmw{12}, 
   \restoregame{gbd12}\mainline{3.Be3} \dtmw{13}, 
   \restoregame{gbd12}\mainline{3.Bxf4} \dtmw{12}, 
   \restoregame{gbd12}\mainline{3.Qe3} \dtmw{14}\color{black}) 
 
   \begin{itemize}
     \item  \varid{1}{2.1} \restoregame{gbd12}\mainline{3.bxc4 Rxb2}
       \draw is a tricky draw in which White appear to be losing but 
       can hold. The mainline is the following \mainline{4.cxd5 Qxd5
         5.exd5 Nd4 6.Nxd4 cxd4} at this point Black will regain the
       Queen and the bishop by force (otherwise White get mated) and
       end up in a ending like this one \mainline{7.Bb4 Rxe2+ 8. Kxe2
         Bxb4} and the position is a curious draw (clearly White
       cannot win which is enough for our oracle). 
 
     \item  \varid{1}{2.2} \restoregame{gbd12}\mainline{3.cxd5 Qxd5} 
     \storegame{gbd12a} 
     (\color{red}\restoregame{gbd12a}\mainline{4.b4} \dtmw{11},
      \restoregame{gbd12a}\mainline{4.d4} \dtmw{10},
      \restoregame{gbd12a}\mainline{4.Nb4} \dtmw{9},
      \restoregame{gbd12a}\mainline{4.Nd4} \dtmw{8},
      \restoregame{gbd12a}\mainline{4.Ne3} \dtmw{11},
      \restoregame{gbd12a}\mainline{4.Bb4} \dtmw{7},
      \restoregame{gbd12a}\mainline{4.Bc3} \dtmw{11},
      \restoregame{gbd12a}\mainline{4.Be3} \dtmw{7},
      \restoregame{gbd12a}\mainline{4.Bxf4} \dtmw{9},
      \restoregame{gbd12a}\mainline{4.Qe3} \dtmw{11},
      \restoregame{gbd12a}\mainline{4.Qe4} \dtmw{11},
      \restoregame{gbd12a}\mainline{4.Qxe5+} \dtmw{3}\color{black})

    \begin{itemize}
      \item \varid{1}{2.2.1}\restoregame{gbd12a}\mainline{4.bxc4
          Rxb2} \draw see line \varid{1}{2.1}.
      \item \varid{1}{2.2.2}\restoregame{gbd12a}\mainline{4.dxc4 Qe6}
          \wloss since White is restricted by Black pawns that
          completly control the dark squares and cannot move his
          Knight, hence his Rook. Black may just move his Queen
          between e6 f5. \justif (\color{red}\mainline{5.b4 cxb4} \dtmw{22}\color{black}). 
    \end{itemize}
  \end{itemize}

  \item \varid{1}{3} \restoregame{gbd1}\mainline{2.Ne3 fxe3+}
        \storegame{gbd1a}
        (\color{red}
        \mainline{3.Qxe3} \dtmw{25}, 
         \restoregame{gbd1a}\mainline{3.Kxe3} \dtmw{22}\color{black}) 

         \restoregame{gbd1a} \mainline{3.Bxe3 d4} \wloss 
         White cannot win \justif \mainline{4.Bc2 dxc3 5.Bxc3 b4
           6.Bd2 Nd4 7.Qe3 Rc6} White is in zugzwang and must give
         another piece. 
  
  \item \varid{1}{4} \restoregame{gbd1}\mainline{2.Be3 d4}
        \storegame{gbd17} 
        (\color{red}\restoregame{gbd17}\mainline{3.Nb4 fxe3+} \dtmw{9}, 
         \restoregame{gbd17}\mainline{3.Nxd4 fxe3+} \dtmw{17}, 
         \restoregame{gbd17}\mainline{3.Bd2} \dtmw{25},
         \restoregame{gbd17}\mainline{3.Bxd4 cxd4}\dtmw{13}, 
         \restoregame{gbd17}\mainline{3.Bxf4 dxc3} \dtmw{16}, 
         \restoregame{gbd17}\mainline{3.Qd2} \dtmw{28}
          \restoregame{gbd17}\mainline{3.b4 dxc3} 
         \dtmw{11} Black either promotes c pawn or is a Rook up (and 
          \mainline{4.Rb2 Qxb2 5.Bxc5 Nd4} is not helping)\color{black})
  \begin{itemize}
    \item \varid{1}{4.1} \restoregame{gbd17}\mainline{3.c4 fxe3+}
      \storegame{gbd172} (\color{red}\mainline{4.Qxe3} \dtmw{13}\color{black}) 

      \restoregame{gbd172}
      \mainline{4.Nxe3 dxe3+} \draw because on each recapture by White
      Black closes the position with \bmove{b4} and white cannot break
      through since \wmove{f4} leads to a quick defeat.

    \item \varid{1}{4.2} \restoregame{gbd17}\mainline{3.cxd4 cxd4}
          \draw \storegame{gbd173} 
         (\color{red}\restoregame{gbd173}\mainline{4.b4} \dtmw{17},
          \restoregame{gbd173}\mainline{4.Nb4} \dtmw{9},
          \restoregame{gbd173}\mainline{4.Nxd4} \dtmw{14}, 
          \restoregame{gbd173}\mainline{4.Bd2} \dtmw{14}, 
          \restoregame{gbd173}\mainline{4.Bxd4 Nxd4} \dtmw{12}, 
          \restoregame{gbd173}\mainline{4.Qd2} \dtmw{15}\color{black})
           \justif \restoregame{gbd173}\mainline{4.Bxf4 exf4 5.b4}
          (otherwise White is a piece down and will lose)
          \mainline{5... Be5} and the position is completely blocked on the
          dark squares. 

  \end{itemize}
\end{itemize}
\restoregame{gbd1}\mainline{2.exd5 Qxd5} \storegame{gdb14} 
 (\color{red}\restoregame{gdb14}\mainline{3.d4 exd4} \dtmw{21},
 \restoregame{gdb14}\mainline{3.Nb4 cxb4} \dtmw{11},
 \restoregame{gdb14}\mainline{3.Nd4 exd4} \dtmw{16},
 \restoregame{gdb14}\mainline{3.Ne3 fxe3+} \dtmw{52},
 \restoregame{gdb14}\mainline{3.Be3 fxe3+} \dtmw{44},
 \restoregame{gdb14}\mainline{3.Bxf4 exf4} \dtmw{15},
 \restoregame{gdb14}\mainline{3.Qe3 fxe3} \dtmw{10},
 \restoregame{gdb14}\mainline{3.Qxe5+ Nxe5} \dtmw{2}\color{black})
\begin{itemize}
    \item \varid{2}{1} \restoregame{gdb14}\mainline{3.b4 cxb4}
         \storegame{gbd1421} 
        (\color{red}\restoregame{gbd1421}\mainline{4.c4} \dtmw{16},
        \restoregame{gbd1421}\mainline{4.d4} \dtmw{15},
        \restoregame{gbd1421}\mainline{4.Rb3} \dtmw{12},
        \restoregame{gbd1421}\mainline{4.Rxb4} \dtmw{14},
        \restoregame{gbd1421}\mainline{4.Nxb4} \dtmw{13},
        \restoregame{gbd1421}\mainline{4.Nd4} \dtmw{12},
        \restoregame{gbd1421}\mainline{4.Ne3} \dtmw{14},
        \restoregame{gbd1421}\mainline{4.Be3} \dtmw{10},
        \restoregame{gbd1421}\mainline{4.Bxf4} \dtmw{10},
        \restoregame{gbd1421}\mainline{4.Qe3} \dtmw{9},
        \restoregame{gbd1421}\mainline{4.Qe4} \dtmw{14},
        \restoregame{gbd1421}\mainline{4.Qxe5+} \dtmw{5}\color{black})

        \restoregame{gbd1421}\mainline{4.cxb4 Nd4}
          \storegame{gbd1421a}
          (\color{red}\restoregame{gbd1421a}\mainline{5.Rb3}\dtmw{7},
          \restoregame{gbd1421a}\mainline{5.Nxd4}\dtmw{8},
          \restoregame{gbd1421a}\mainline{5.Ne3}\dtmw{14}, 
          \restoregame{gbd1421a}\mainline{5.Bc3}\dtmw{9},
          \restoregame{gbd1421a}\mainline{5.Be3}\dtmw{9},
          \restoregame{gbd1421a}\mainline{5.Qe3}\dtmw{2},
          \restoregame{gbd1421a}\mainline{5.Qxe5+}\dtmw{5},
          \restoregame{gbd1421a}\mainline{5.Bxf4}\dtmw{8} \color{black}) 

          \restoregame{gbd1421a}\mainline{5.Qe4 Qxe4} \draw
          \storegame{gbd1421b}
          (\color{red}\restoregame{gbd1421b}\mainline{6.Rb3 Qe2} checkmate,
          \restoregame{gbd1421b}\mainline{6.Nxd4} \dtmw{5},
          \restoregame{gbd1421b}\mainline{6.Ne3 Qxf3} checkmate,
          \restoregame{gbd1421b}\mainline{6.Bc3 Qxf3} checkmate,
          \restoregame{gbd1421b}\mainline{6.Be3 Qxf3} checkmate,
          \restoregame{gbd1421b}\mainline{6.Bxf4 Qxf3} checkmate\color{black})

          on either d-pawn or f-pawn capture of the Queen Black 
          plays \bmove{Nxc2} and then his King on e6-f6 squares. The
          position is completly blocked. 

    \item \varid{2}{2} \restoregame{gdb14}\mainline{3.c4 bxc4}
          by transposition we have reached line \varid{1}{2.2}. 

  \end{itemize} 
\restoregame{gdb14}\mainline{3.Qe4 Qxe4}
\storegame{gbd1410}
           (\color{red}\restoregame{gbd1410}\mainline{4.b4} \dtmw{7},
            \restoregame{gbd1410}\mainline{4.c4} \dtmw{6},
            \restoregame{gbd1410}\mainline{4.d4} \dtmw{7},
            \restoregame{gbd1410}\mainline{4.Nb4} \dtmw{10},
            \restoregame{gbd1410}\mainline{4.Nd4} \dtmw{6},
            \restoregame{gbd1410}\mainline{4.Ne3} \dtmw{7},
            \restoregame{gbd1410}\mainline{4.Be3 Qxe3} checkmate,
            \restoregame{gbd1410}\mainline{4.Bxf4} \dtmw{8}\color{black}) 
\begin{itemize}
  \item \varid{3}{1}\restoregame{gbd1410} \mainline{4.dxe4 b4}
        \storegame{gbd31}
        (\color{red}\restoregame{gbd31}\mainline{5.Nxb4}\dtmw{27},
        \restoregame{gbd31}\mainline{5.Nd4}\dtmw{9},
        \restoregame{gbd31}\mainline{5.Bxf4}\dtmw{7},
        \restoregame{gbd31}\mainline{5.Be3}\dtmw{6}\color{black})

  \begin{itemize}
    \item \varid{3}{1.1} \restoregame{gbd31} \mainline{5.c4 Ke6}
           \draw the position is completely blocked and Black can 
          just move his King on e6-f6.

    \item \varid{3}{1.2} \restoregame{gbd31} \mainline{5.cxb4 cxb4}
         \draw White pieces are blocked, his only active plan is to 
         bring the King on c4 but Black can play its bishop on 
         c5-f2 \justif \mainline{6.Ke2 Bc5 7.Kd3 Ke6 8.Kc4 Kd6}. 
  
    \item \varid{3}{1.3} \restoregame{gbd31} \mainline{5.Ne3 
          fxe3+} \storegame{gbd313} 
          (\color{red}\restoregame{gbd313}\mainline{6.Bxe3} \dtmw{10},
           \restoregame{gbd313}\mainline{6.Ke2} \dtmw{14}\color{black})
           
           \restoregame{gbd313}\mainline{6.Kxe3 bxc3} \storegame{gbd313a}
           (\color{red}\restoregame{gbd313a} \mainline{7.b4} \dtmw{5},
           \restoregame{gbd313a} \mainline{7.Rc2} \dtmw{11},
           \restoregame{gbd313a} \mainline{7.Kd3} \dtmw{5},
           \restoregame{gbd313a} \mainline{7.Ke2} \dtmw{5},
           \restoregame{gbd313a} \mainline{7.Kf2} \dtmw{5},
           \restoregame{gbd313a} \mainline{7.f4} \dtmw{5}\color{black})

           \restoregame{gbd313a} \mainline{7.Bxc3 c4} \storegame{gbd313b}
           (\color{red}\restoregame{gbd313b}\mainline{8.b4}\dtmw{23},
           \restoregame{gbd313b}\mainline{8.Ke2}\dtmw{15},
           \restoregame{gbd313b}\mainline{8.Kd2}\dtmw{28},
           \restoregame{gbd313b}\mainline{8.Kf2}\dtmw{13},
           \restoregame{gbd313b}\mainline{8.f4}\dtmw{25},
           \restoregame{gbd313b}\mainline{8.Bb4}\dtmw{9},
           \restoregame{gbd313b}\mainline{8.Bd4}\dtmw{7},
           \restoregame{gbd313b}\mainline{8.Bxe5+}\dtmw{8},
           \restoregame{gbd313b}\mainline{8.Bd2}\dtmw{9},
           \restoregame{gbd313b}\mainline{8.Rc2}\dtmw{10},
           \restoregame{gbd313b}\mainline{8.Rd2}\dtmw{10},
           \restoregame{gbd313b}\mainline{8.Re2}\dtmw{8},
           \restoregame{gbd313b}\mainline{8.Rf2}\dtmw{8}\color{black})

           \restoregame{gbd313b}\mainline{8.bxc4 Rxb2} \wloss
           \justif \mainline{9.Bxb2 Ke6} white cannot win since after
           \bmove{Bc5} White clearly cannot progress.            

    \item \varid{3}{1.4} \restoregame{gbd31} \mainline{5.Ke2 bxc3}
           \storegame{gbd314}
          (\color{red}\restoregame{gbd314} \mainline{6.Kd3} \dtmw{5},
           \restoregame{gbd314} \mainline{6.Bxf4} \dtmw{5},
           \restoregame{gbd314} \mainline{6.Nb4} \dtmw{5},
           \restoregame{gbd314} \mainline{6.Be3} \dtmw{5},
           \restoregame{gbd314} \mainline{6.b4} \dtmw{5},
           \restoregame{gbd314} \mainline{6.Ne3} \dtmw{5},
           \restoregame{gbd314} \mainline{6.Nd4} \dtmw{3},
           \restoregame{gbd314} \mainline{6.Kf2 cxd2=Q+} checkmate\color{black})

          \restoregame{gbd314} \mainline{6.Bxc3 Nd4+} 
          \storegame{gbd314a}
          (\color{red}\restoregame{gbd314a}\mainline{7.Kd2} \dtmw{39},
          \restoregame{gbd314a}\mainline{7.Kf2} \dtmw{16},
          \restoregame{gbd314a}\mainline{7.Nxd4} \dtmw{21}\color{black})
          \begin{itemize}
            \item \varid{3}{1.4.1}\restoregame{gbd314a}\mainline{7.Kd3
              Nxf3} \storegame{gbd3141}
              (\color{red}\restoregame{gbd3141}\mainline{8.Kc4} \dtmw{40},
              \restoregame{gbd3141}\mainline{8.Ke2} \dtmw{21},
              \restoregame{gbd3141}\mainline{8.Bd2} \dtmw{14},
              \restoregame{gbd3141}\mainline{8.Nb4} \dtmw{11},
              \restoregame{gbd3141}\mainline{8.Bb4} \dtmw{13},
              \restoregame{gbd3141}\mainline{8.Nd4} \dtmw{13},
              \restoregame{gbd3141}\mainline{8.Bxe5} \dtmw{10},
              \restoregame{gbd3141}\mainline{8.Bd4} \dtmw{10}\color{black})

             \subitem - \varid{3}{1.4.1.1} \restoregame{gbd3141}
             \mainline{8.b4 Rc6}  \storegame{gbd3141Mehdi}          
               (\color{red}\restoregame{gbd3141Mehdi} \mainline{9.Ke2} \dtmw{41},
             \restoregame{gbd3141Mehdi} \mainline{9.Ke2} \dtmw{24},
                \restoregame{gbd3141Mehdi} \mainline{9.Bd2} \dtmw{14},
                    \restoregame{gbd3141Mehdi} \mainline{9.Kc4} \dtmw{12},
                        \restoregame{gbd3141Mehdi} \mainline{9.Bd4} \dtmw{12},
                            \restoregame{gbd3141Mehdi} \mainline{9.Nd4} \dtmw{11},
                                \restoregame{gbd3141Mehdi} \mainline{9.Bxe5} \dtmw{10},
                           \restoregame{gbd3141Mehdi} \mainline{9.Rb3} \dtmw{9}
                \color{black}) 
                \subsubitem * \varid{3}{1.4.1.1.1}\restoregame{gbd3141Mehdi}
                \mainline{9.b5 Rb6} \draw

                \subsubitem *\varid{3}{1.4.1.1.2}\restoregame{gbd3141Mehdi}
                \mainline{9.bxc5 Bxc5} \draw

                \subsubitem * \varid{3}{1.4.1.1}.3\restoregame{gbd3141Mehdi}
                \mainline{9.Ne3 fxe3} \draw

             \subitem -\varid{3}{1.4.1.2} \restoregame{gbd3141}
             \mainline{8.Ne3 fxe3} \storegame{gbd3141a}
             (\color{red}\restoregame{gbd3141a} \mainline{9.Ke2} \dtmw{41},
             \restoregame{gbd3141a} \mainline{9.Re2} \dtmw{24},
             \restoregame{gbd3141a} \mainline{9.b4} \dtmw{17},
             \restoregame{gbd3141a} \mainline{9.Rc2} \dtmw{15},
             \restoregame{gbd3141a} \mainline{9.Kc4} \dtmw{14},
             \restoregame{gbd3141a} \mainline{9.Bb4} \dtmw{13},
             \restoregame{gbd3141a} \mainline{9.Bxe5} \dtmw{8},
             \restoregame{gbd3141a} \mainline{9.Rd2} \dtmw{7},
             \restoregame{gbd3141a} \mainline{9.Bd4} \dtmw{7},
             \restoregame{gbd3141a} \mainline{9.Bd2} \dtmw{7},
             \restoregame{gbd3141a} \mainline{9.Rf2} \dtmw{5},
             \restoregame{gbd3141a} \mainline{9.Kc2} \dtmw{2}\color{black})

             \restoregame{gbd3141a} \mainline{9.Kxe3 c4} \draw
             since Rook exchanges is unavoidable (otherwise White
             lose) and the Bishop's ending is draw. 

            \item \varid{3}{1.4.2}\restoregame{gbd314a}
                    \mainline{7.Bxd4 exd4}     
                      \storegame{gbd3141Mehdi2}          
               (\color{red}      \restoregame{gbd3141Mehdi2} \mainline{8.Nxd4} \dtmw{18},
               \restoregame{gbd3141Mehdi2} \mainline{8.Kd2} \dtmw{18},
               \restoregame{gbd3141Mehdi2} \mainline{8.e5} \dtmw{13},
               \restoregame{gbd3141Mehdi2} \mainline{8.Nb4} \dtmw{13},
               \restoregame{gbd3141Mehdi2} \mainline{8.Ne3} \dtmw{10},
               \restoregame{gbd3141Mehdi2} \mainline{8.Kf2} \dtmw{8},
                \color{black}) \draw \justif{8.b4 c4 9.Nxd4 Be5} and
                the resulting Rook ending is draw.

            \end{itemize}
        \end{itemize}
\end{itemize}      
\restoregame{gbd1410}\mainline{4.fxe4 b4} 
\storegame{gbd4}
(\color{red}\restoregame{gbd4}\mainline{5.d4} \dtmw{14}, 
\restoregame{gbd4}\mainline{5.Nxb4} \dtmw{21},  
\restoregame{gbd4}\mainline{5.Nd4} \dtmw{13}, 
\restoregame{gbd4}\mainline{5.Be3} \dtmw{7},
\restoregame{gbd4}\mainline{5.Bxf4} \dtmw{7}\color{black}) 
\begin{itemize}
  \item \varid{4}{1} \restoregame{gbd4}\mainline{5.Ke2 Ke6 6.Kf2}
  (or \variation{6.Kf3}) \mainline{6... Kf6} \draw since other moves
  than King loses (see previous lines) or rejoin the mainline
  (\wmove{c4}\color{black}). 

  \item \varid{4}{2} \restoregame{gbd4}\mainline{5.Kf3 Ke6} same as
    line \varid{4}{1}. 
\end{itemize}
\restoregame{gbd4}\mainline{5.c4 Nd4} \draw the position is totally
locked on dark squares and White cannot progress. 
 
   \subsection{White moves \wmove{f4}}
\gardnerstart
\mainline{1.f4 exf4}\storegame{gbe1}
(\color{red}\restoregame{gbe1}\mainline{2.Nb4} \dtmw{20},
\restoregame{gbe1}\mainline{2.Nd4} \dtmw{24},
\restoregame{gbe1}\mainline{2.Qf3}  \dtmw{24},
\restoregame{gbe1}\mainline{2.Kf3 Ne5+} \dtmw{17}, 
\restoregame{gbe1}\mainline{2.c4 Be5} \dtmw{26},
\restoregame{gbe1}\mainline{2.e4 fxe4} \dtmw{18}\color{black})
\begin{itemize}
\item \varid{1}{1} \restoregame{gbe1}\mainline{2.b4 f3}
  \storegame{gbe11} 
  (\color{red}\restoregame{gbe11}\mainline{3.bxc5} \dtmw{14},
   \restoregame{gbe11}\mainline{3.c4} \dtmw{11}, 
   \restoregame{gbe11} \mainline{3.d4} \dtmw{11}, 
   \restoregame{gbe11}\mainline{3.e4} \dtmw{9}, 
   \restoregame{gbe11}\mainline{3.Rb3} \dtmw{9},
   \restoregame{gbe11}\mainline{3.Nd4} \dtmw{11}, 
   \restoregame{gbe11} \mainline{3.Kxf3} \dtmw{24}\color{black})

  \restoregame{gbe11}
  \mainline{3.Qxf3 Ne5} \storegame{gbe11c}
  (\color{red}\restoregame{gbe11c}\mainline{4.c4} \dtmw{14},
    \restoregame{gbe11c}\mainline{4.d4} \dtmw{14},
    \restoregame{gbe11c}\mainline{4.e4} \dtmw{11}
    \restoregame{gbe11c}\mainline{4.Rb3} \dtmw{13},
    \restoregame{gbe11c}\mainline{4.Nd4} \dtmw{19},
    \restoregame{gbe11c}\mainline{4.Qe2} \dtmw{23},
    \restoregame{gbe11c}\mainline{4.Qf4} \dtmw{9},
    \restoregame{gbe11c}\mainline{4.Qxf5+} \dtmw{2},
    \restoregame{gbe11c}\mainline{4.Qe4} \dtmw{7},
    \restoregame{gbe11c}\mainline{4.Qxd5} \dtmw{6},
    \restoregame{gbe11c}\mainline{4.Ke2} \dtmw{20}\color{black}) 

    \restoregame{gbe11c}\mainline{4.bxc5 Nxd3+ 5.Ke2 Nxc5}
    \storegame{gbe11d}
    (\color{red}\restoregame{gbe11d}\mainline{6.Nb4} \dtmw{37} 
    \restoregame{gbe11d}\mainline{6.c4} \dtmw{14},
    \restoregame{gbe11d}\mainline{6.e4} \dtmw{11},
    \restoregame{gbe11d}\mainline{6.Rb3} \dtmw{12},
    \restoregame{gbe11d}\mainline{6.Rxb5} \dtmw{14},
     \restoregame{gbe11d}\mainline{6.Qf2} \dtmw{21},
     \restoregame{gbe11d}\mainline{6.Qf4} \dtmw{12},
     \restoregame{gbe11d}\mainline{6.Qe4} \dtmw{8},
     \restoregame{gbe11d}\mainline{6.Qxd5} \dtmw{10},
     \restoregame{gbe11d}\mainline{6.Qxf5+} \dtmw{9}\color{black})

    \begin{itemize}
      \item \varid{1}{1.1} \restoregame{gbe11d}\mainline{6.Rb4 Ne4}
            \storegame{gbe111} 
           (\color{red}\restoregame{gbe111}\mainline{7.c4}  \dtmw{12},
            \restoregame{gbe111}\mainline{7.Rb3} \dtmw{21}, 
            \restoregame{gbe111}\mainline{7.Rb2} \dtmw{37}, 
            \restoregame{gbe111}\mainline{7.Kd3} \dtmw{29}, 
            \restoregame{gbe111}\mainline{7.Qf2} \dtmw{13},
            \restoregame{gbe111}\mainline{7.Qf4} \dtmw{15}\color{black}) 

            \restoregame{gbe111}\mainline{7.Nd4 Nxd2 8.Kxd2 Qe4} \draw
            \storegame{gbe111a} \justif \mainline{9.Qxe4 fxe4 10.Rxb5 Rxb5
            11.Nxb5} and the ending Knight vs. Bishop is draw or 
            \restoregame{gbe111a}
            \justif \mainline{9.Rxb5 Qxf3 10.Rxb6 Qf2+ 11.Kd3 Ke6 12.Nc6+} and
            perpetual check. 

          \item \varid{1}{1.2} \restoregame{gbe11d} \mainline{6.Nd4
              Qe4} Because of the threat to the White king, White's
            move is forced \mainline{7.Qxe4 fxe4} \wloss White is
            blocked and must take the b pawn with his Rook if he looks
            for any progress after the rook exchanges Black easily
            draw. If \mainline{8.Nxb4 Nc2 9.Rb3 Bc5} White is in
            zugzwang and lose.
    \end{itemize}
    
  \item \varid{1}{3} \restoregame{gbe1}\mainline{2.d4 fxe3+}
     \storegame{gbe13}
    (\color{red}\restoregame{gbe13}\mainline{3.Nxe3 cxd4} \dtmw{26}, 
     \restoregame{gbe13}\mainline{3.Qxe3} \dtmw{41}, 
     \restoregame{gbe13}\mainline{3.Kf3 Qe4} checkmate)

      \restoregame{gbe13}\mainline{3.Bxe3 Qe4}
      \storegame{gbe132}      
     (\color{red}\restoregame{gbe132}\mainline{4.Qf3} \dtmw{48},
      \restoregame{gbe132}\mainline{4.Qd2} \dtmw{30},
      \restoregame{gbe132}\mainline{4.b4} \dtmw{35},
      \restoregame{gbe132}\mainline{4.c4} \dtmw{16},
      \restoregame{gbe132}\mainline{4.Nb4} \dtmw{15},
      \restoregame{gbe132}\mainline{4.Bd2} \dtmw{24},
      \restoregame{gbe132}\mainline{4.Bf4} \dtmw{18},
      \restoregame{gbe132}\mainline{4.Qd3} \dtmw{9},
      \restoregame{gbe132}\mainline{4.Qc4} \dtmw{8},
      \restoregame{gbe132}\mainline{4.Qxb5} \dtmw{8}\color{black}) 

      \restoregame{gbe132} \mainline{4.dxc5 Bxc5} 
      \storegame{gbe132a} 
      (\color{red}\restoregame{gbe132a}\mainline{5.c4} \dtmw{4},
      \restoregame{gbe132a}\mainline{5.Nb4} \dtmw{14},
      \restoregame{gbe132a}\mainline{5.Nd4} \dtmw{14}
      \restoregame{gbe132a}\mainline{5.Qd2} \dtmw{5},
      \restoregame{gbe132a}\mainline{5.Qd3} \dtmw{5},
      \restoregame{gbe132a}\mainline{5.Qc4} \dtmw{5},
      \restoregame{gbe132a}\mainline{5.Qxb5} \dtmw{5},
      \restoregame{gbe132a}\mainline{5.Qf3} \dtmw{32}, 
      \restoregame{gbe132a}\mainline{5.Bc4} \dtmw{15}\color{black})
      \begin{itemize}
        \item \varid{1}{3.1} \restoregame{gbe132a}\mainline{5.b4
          Bd6} the only non losing move is \mainline{6.Bxb6 Qf4+}
          \draw perpertual check on d2 f4.

        \item  \varid{1}{3.2} \restoregame{gbe132a}\mainline{5.Bxc5
        Qf4+} \draw perpertual check on d2 f4. 
      \end{itemize}
\end{itemize}
\restoregame{gbe1}\mainline{2.exf4 Qxe2+ 3.Kxe2 d4} \storegame{gbe3}
(\color{red}\restoregame{gbe3}\mainline{4.Nb4 cxb4}  \dtmw{24},
 \restoregame{gbe3}\mainline{4.Nxd4 cxd4} \dtmw{31}, 
 \restoregame{gbe3}\mainline{4.Ne3 dxe3} \dtmw{31}, 
 \restoregame{gbe3}\mainline{4.Be3 dxe3} \dtmw{8}\color{black})
\begin{itemize}
  \item \varid{4}{1} \restoregame{gbe3}\mainline{4.b4 Ke6} 
         \storegame{gbe31}
         (\color{red}\restoregame{gbe31}\mainline{5.Ne3} \dtmw{32},
         \restoregame{gbe31} \mainline{5.Be3} \dtmw{40},
         \restoregame{gbe31} \mainline{5.Nxd4} \dtmw{28}\color{black})
  \begin{itemize}
    \item \varid{4}{1.1} \restoregame{gbe31}
          \mainline{5.bxc5 Bxc5} \storegame{gbe311}
           (\color{red}\restoregame{gbe311}\mainline{6.Kf3} \dtmw{23},
           \restoregame{gbe311}\mainline{6.Rb4} \dtmw{23},
           \restoregame{gbe311}\mainline{6.Ne3} \dtmw{22},
           \restoregame{gbe311}\mainline{6.Rxb5} \dtmw{12},
           \restoregame{gbe311}\mainline{6.Kf2} \dtmw{7}\color{black})
    \begin{itemize}
      \item \restoregame{gbe311} \varid{4}{1.1.1} \mainline{6.c4 bxc4}
             \storegame{gbe3111}
             (\color{red}\restoregame{gbe3111}\mainline{7.Rb3} \dtmw{6},
             \restoregame{gbe3111}\mainline{7.Rb4} \dtmw{10},
             \restoregame{gbe3111}\mainline{7.Rb5} \dtmw{8},
             \restoregame{gbe3111}\mainline{7.Nb4} \dtmw{9},
             \restoregame{gbe3111}\mainline{7.Nxd4} \dtmw{6},
             \restoregame{gbe3111}\mainline{7.Ne3} \dtmw{10},
             \restoregame{gbe3111}\mainline{7.dxc4} \dtmw{8},
             \restoregame{gbe3111}\mainline{7.Bc3} \dtmw{12},
             \restoregame{gbe3111}\mainline{7.Bb4} \dtmw{10},
             \restoregame{gbe3111}\mainline{7.Be3} \dtmw{7},
             \restoregame{gbe3111}\mainline{7.Kf2} \dtmw{6},
             \restoregame{gbe3111}\mainline{7.Kf3} \dtmw{7}\color{black})

             \restoregame{gbe3111} \mainline{7.Rxb6 Bxb6}
             \storegame{gbe3111a}
             (\color{red}\restoregame{gbe3111a}\mainline{8.Nb4} \dtmw{11},
             \restoregame{gbe3111a} \mainline{8.Nxd4} \dtmw{13},
             \restoregame{gbe3111a} \mainline{8.Ne3} \dtmw{14},
             \restoregame{gbe3111a} \mainline{8.Bb4} \dtmw{15},
             \restoregame{gbe3111a} \mainline{8.Bc3} \dtmw{14},
             \restoregame{gbe3111a} \mainline{8.Be3} \dtmw{14},
             \restoregame{gbe3111a} \mainline{8.Kf2} \dtmw{10},
             \restoregame{gbe3111a} \mainline{8.Kf3} \dtmw{10}\color{black})

             \restoregame{gbe3111a} \mainline{8.dxc4 
              Bc5} \wloss White is blocked and cannot untangle if
              Black just moves his King to d6-e6.

      \item \restoregame{gbe311} \varid{4}{1.1.2} \mainline{6.cxd4
          Nxd4} \storegame{gbe4112}
         (\color{red} \restoregame{gbe4112}\mainline{7.Kf2}\dtmw{13},
         \restoregame{gbe4112}\mainline{7.Ke3}\dtmw{13}\color{black})
          \restoregame{gbe4112} \mainline{7.Nxd4+ Bxd4} \draw
          the Black King blocks the position on d5.
    
      \item \restoregame{gbe311} \varid{4}{1.1.3} 
              \mainline{6.Rb3 dxc3}\storegame{gbe4113} 
              (\color{red}\restoregame{gbe4113}\mainline{7.Rxb5} \dtmw{11},
              \restoregame{gbe4113}\mainline{7.d4} \dtmw{15},
              \restoregame{gbe4113}\mainline{7.Rb4} \dtmw{11},
              \restoregame{gbe4113}\mainline{7.Rb2} \dtmw{5},
              \restoregame{gbe4113}\mainline{7.Nb4} \dtmw{4},
              \restoregame{gbe4113}\mainline{7.Nd4+} \dtmw{4},
              \restoregame{gbe4113}\mainline{7.Ne3} \dtmw{2},
              \restoregame{gbe4113}\mainline{7.Kf3}
              \dtmw{4}\color{black}) 

              \subitem - \varid{4}{1.1.3.1}
              \restoregame{gbe4113}\mainline{7.Bxc3 b4} 
               Black moves his King to d5 and blocks the position.

              \subitem - \varid{4}{1.1.3.2}
              \restoregame{gbe4113}\mainline{7.Rxc3 Nd4+}
              \storegame{gbe4113a}
              (\color{red} \restoregame{gbe4113a}\mainline{8.Ke3}
                \dtmw{11}\color{black})
              
              \subsubitem * \restoregame{gbe4113a} \mainline{8.Nxd4+
                Bxd4} \draw Black King comes to d5 and blocks the 
                position.

             \subsubitem * \restoregame{gbe4113a} \mainline{8.Kf2 Kd5} 
                \draw
              
              \subitem - \varid{4}{1.1.3.3}
              \restoregame{gbe4113}\mainline{7.Be3 Bxe3} \draw
              after \bmove{b4} and \bmove{Kd5} Black locks down the
              position and White cannot progress.

      \item \restoregame{gbe311} \varid{4}{1.1.4} 
             \mainline{6.Nb4 dxc3} \storegame{gbe3114}
             (\color{red}\restoregame{gbe3114}\mainline{7.Rc2} \dtmw{8},
              \restoregame{gbe3114}\mainline{7.Nxc6} \dtmw{7},
              \restoregame{gbe3114}\mainline{7.d4} \dtmw{6},
               \restoregame{gbe3114}\mainline{7.Kf3} \dtmw{5}, 
               \restoregame{gbe3114}\mainline{7.Rb3} \dtmw{4},
               \restoregame{gbe3114}\mainline{7.Nd5} \dtmw{5},
               \restoregame{gbe3114}\mainline{7.Nc2}\dtmw{5}\color{black}) 

               \restoregame{gbe3114}\mainline{7.Bxc3 Nxb4}
               \storegame{gbe3114a}
                 (\color{red}\restoregame{gbe3114a}\mainline{8.Rxb4} \dtmw{17},
                 \restoregame{gbe3114a}\mainline{8.Ke2} \dtmw{17},
                 \restoregame{gbe3114a}\mainline{8.Rd2} \dtmw{16},
                 \restoregame{gbe3114a}\mainline{8.Be5} \dtmw{16},
                 \restoregame{gbe3114a}\mainline{8.Rb3} \dtmw{15},
                 \restoregame{gbe3114a}\mainline{8.Bd2} \dtmw{14},
                 \restoregame{gbe3114a}\mainline{8.Kf3} \dtmw{14},
                 \restoregame{gbe3114a}\mainline{8.Bf6} \dtmw{13},
                 \restoregame{gbe3114a}\mainline{8.Bd4} \dtmw{11},
                 \restoregame{gbe3114a}\mainline{8.Rc3} \dtmw{9}\color{black}) 

               \subitem \varid{4}{1.1.4.1}\restoregame{gbe3114a} 
               \mainline{8.Bxb4 Bxb4} 
               \draw this Rook ending is clearly draw. 

               \subitem \varid{4}{1.1.4.2}\restoregame{gbe3114a} 
               \mainline{8.d4 Bd6} \storegame{gbe3114b}
               (\color{red}\restoregame{gbe3114b} \mainline{9.d5+} \dtmw{18},
               \restoregame{gbe3114b} \mainline{9.Rb3} \dtmw{16},
               \restoregame{gbe3114b} \mainline{9.Rxb4} \dtmw{15},
               \restoregame{gbe3114b} \mainline{9.Rc2} \dtmw{10},
               \restoregame{gbe3114b} \mainline{9.Rd2} \dtmw{14},
               \restoregame{gbe3114b} \mainline{9.Bd2} \dtmw{18},
               \restoregame{gbe3114b} \mainline{9.Kd2} \dtmw{18},
               \restoregame{gbe3114b} \mainline{9.Ke3} \dtmw{16},
               \restoregame{gbe3114b} \mainline{9.Kf3} \dtmw{19},
               \restoregame{gbe3114b} \mainline{9.Kf2} \dtmw{12}\color{black}) 

               \restoregame{gbe3114b} \mainline{9.Bxb4 Bxb4} \draw
               this is the same ending as variation \varid{1}{1.1.4.1}. 

      \item \restoregame{gbe311} \varid{4}{1.1.5} \mainline{6.Nxd4+
          Nxd4} \storegame{gbe3115}
         (\color{red}\restoregame{gbe3115}\mainline{7.Ke3} \dtmw{23},
          \restoregame{gbe3115}\mainline{7.Kf2} \dtmw{25}\color{black}) 

          \restoregame{gbe3115}
          \mainline{7.cxd4 Bxd4} \storegame{gbe3115a}
           (\color{red}\restoregame{gbe3115a}\mainline{8.Rc2} \dtmw{20},
           \restoregame{gbe3115a}\mainline{8.Rxb5} \dtmw{10},
           \restoregame{gbe3115a}\mainline{8.Bc3} \dtmw{11},
           \restoregame{gbe3115a}\mainline{8.Bb4} \dtmw{11},
           \restoregame{gbe3115a}\mainline{8.Be3} \dtmw{21},
           \restoregame{gbe3115a}\mainline{8.Kf3} \dtmw{10}\color{black}) \draw
           since Black King may move to seat on d5 and block the
           position. If Bishops are exchanged the resulting Rook
           ending is clearly draw.

      \item \restoregame{gbe311} \varid{4}{1.1.6} 
             \mainline{6.Be3 dxc3} \storegame{gbe3116}
             (\color{red}\restoregame{gbe3116}\mainline{7.d4} \dtmw{9},
             \restoregame{gbe3116} \mainline{7.Rb4} \dtmw{10},
             \restoregame{gbe3116} \mainline{7.Rxb5} \dtmw{12},
             \restoregame{gbe3116} \mainline{7.Nb4} \dtmw{5},
             \restoregame{gbe3116} \mainline{7.Nd4+} \dtmw{9},
             \restoregame{gbe3116} \mainline{7.Kf2} \dtmw{7},
             \restoregame{gbe3116} \mainline{7.Kf3} \dtmw{7},
             \restoregame{gbe3116} \mainline{7.Bd2} \dtmw{5},
             \restoregame{gbe3116} \mainline{7.Bf2} \dtmw{7},
             \restoregame{gbe3116} \mainline{7.Bd4} \dtmw{9},
             \restoregame{gbe3116} \mainline{7.Bxc5} \dtmw{7}\color{black})

             \restoregame{gbe3116} \mainline{7.Rb3 Bxe3}
             \storegame{gbe3116a}
             (\color{red}\restoregame{gbe3116a}\mainline{8.d4} \dtmw{11},
             \restoregame{gbe3116a} \mainline{8.Rb2} \dtmw{5},
             \restoregame{gbe3116a} \mainline{8.Rb4} \dtmw{9},
             \restoregame{gbe3116a} \mainline{8.Rxb5} \dtmw{7},
             \restoregame{gbe3116a} \mainline{8.Nb4} \dtmw{10},
             \restoregame{gbe3116a} \mainline{8.Nd4+} \dtmw{5},
             \restoregame{gbe3116a} \mainline{8.Nxe3} \dtmw{8},
             \restoregame{gbe3116a} \mainline{8.Kf3} \dtmw{9},
             \restoregame{gbe3116a} \mainline{8.Rxc3} \dtmw{13}\color{black})

             \restoregame{gbe3116a} \mainline{8.Kxe3 b4}
             \storegame{gbe3116b}
             (\color{red}\restoregame{gbe3116b}\mainline{9.Rxb4} \dtmw{7},
             \restoregame{gbe3116b} \mainline{9.Rxc3} \dtmw{9},
             \restoregame{gbe3116b} \mainline{9.Rb2} \dtmw{5},
             \restoregame{gbe3116b} \mainline{9.Nxb4} \dtmw{11},
             \restoregame{gbe3116b} \mainline{9.Nd4} \dtmw{6},
             \restoregame{gbe3116b} \mainline{9.Ke2} \dtmw{15},
             \restoregame{gbe3116b} \mainline{9.Kf2} \dtmw{12},
             \restoregame{gbe3116b} \mainline{9.Kf3} \dtmw{18}\color{black})

             \restoregame{gbe3116b} \mainline{9.d4 Kd5}
             \storegame{gbe3116c}
             (\color{red}\restoregame{gbe3116c}\mainline{10.Rxb4} \dtmw{5},
             \restoregame{gbe3116c} \mainline{10.Rxc3} \dtmw{8},
             \restoregame{gbe3116c} \mainline{10.Rb2} \dtmw{4},
             \restoregame{gbe3116c} \mainline{10.Nxb4+} \dtmw{14},
             \restoregame{gbe3116c} \mainline{10.Ke2} \dtmw{9},
             \restoregame{gbe3116c} \mainline{10.Kf2} \dtmw{9},
             \restoregame{gbe3116c} \mainline{10.Kf3} \dtmw{8}\color{black})

             \restoregame{gbe3116c} \mainline{10.Kd3 Rb5}
             \storegame{gbe3116d}
             (\color{red}\restoregame{gbe3116d}\mainline{11.Rxb4} \dtmw{8},
             \restoregame{gbe3116d} \mainline{11.Rxc3} \dtmw{17},
             \restoregame{gbe3116d} \mainline{11.Rb2} \dtmw{6},
             \restoregame{gbe3116d} \mainline{11.Ke2} \dtmw{6},
             \restoregame{gbe3116d} \mainline{11.Nxb4+} \dtmw{15},
             \restoregame{gbe3116d} \mainline{11.Ke3} \dtmw{8}\color{black})

             \restoregame{gbe3116d} \mainline{11.Ne3+ Kd6}
             \storegame{gbe3116e}
             (\color{red}\restoregame{gbe3116e}\mainline{12.Rb2} \dtmw{8},
             \restoregame{gbe3116e} \mainline{12.Rxb4} \dtmw{10},
             \restoregame{gbe3116e} \mainline{12.Rxc3} \dtmw{16},
             \restoregame{gbe3116e} \mainline{12.Kc2} \dtmw{10},
             \restoregame{gbe3116e} \mainline{12.Ke2} \dtmw{9},
             \restoregame{gbe3116e} \mainline{12.Kc4} \dtmw{16},
             \restoregame{gbe3116e} \mainline{12.Nc4+} \dtmw{19},
             \restoregame{gbe3116e} \mainline{12.Nd5} \dtmw{7},
             \restoregame{gbe3116e} \mainline{12.Nxf5+} \dtmw{9},
             \restoregame{gbe3116e} \mainline{12.d5} \dtmw{14}\color{black})
 
            \restoregame{gbe3116e} \mainline{12.Nc2 Kd5} \draw
             by repetition. 
    \end{itemize}

    \item \varid{4}{1.2} \restoregame{gbe31}\mainline{5.c4 bxc4} 
          \storegame{gb312} 
         (\color{red}\restoregame{gb312}\mainline{6.bxc6} \dtmw{10},
          \restoregame{gb312}\mainline{6.b5} \dtmw{19},
          \restoregame{gb312}\mainline{6.Rb3} \dtmw{7},
          \restoregame{gb312}\mainline{6.Nxd4} \dtmw{},
          \restoregame{gb312}\mainline{6.Ne3} \dtmw{12},
          \restoregame{gb312}\mainline{6.Bc3} \dtmw{8},
          \restoregame{gb312}\mainline{6.Be3} \dtmw{11},
          \restoregame{gb312}\mainline{6.Kf2} \dtmw{12},
          \restoregame{gb312}\mainline{6.Kf3} \dtmw{11}\color{black})
        
         \restoregame{gb312}\mainline{6.dxc4 cxb4} 
         \storegame{gbe312a}
         (\color{red}\restoregame{gbe312a}\mainline{7.c5} \dtmw{20},
         \restoregame{gbe312a}\mainline{7.Rxb4} \dtmw{12},
         \restoregame{gbe312a}\mainline{7.Nxb4} \dtmw{24},
         \restoregame{gbe312a}\mainline{7.Nxd4+} \dtmw{22},
         \restoregame{gbe312a}\mainline{7.Bc3} \dtmw{12},
         \restoregame{gbe312a}\mainline{7.Bxb4} \dtmw{20},
         \restoregame{gbe312a}\mainline{7.Be3} \dtmw{16},
         \restoregame{gbe312a}\mainline{7.Kf3} \dtmw{21},
         \restoregame{gbe312a}\mainline{7.Kf2} \dtmw{22}\color{black})
    \begin{itemize}
      \item \varid{4}{1.2.1} \restoregame{gbe312a} 
            \mainline{7.Ne3 dxe3}\storegame{gbe4121} 
            (\color{red}\restoregame{gbe4121} \mainline{8.Rb3} \dtmw{8},
            \restoregame{gbe4121} \mainline{8.Rxb4} \dtmw{7},
            \restoregame{gbe4121} \mainline{8.Rc2} \dtmw{10},
            \restoregame{gbe4121} \mainline{8.Bc3} \dtmw{7},
            \restoregame{gbe4121} \mainline{8.Bxb4} \dtmw{10},
            \restoregame{gbe4121} \mainline{8.Kd3} \dtmw{11},
            \restoregame{gbe4121} \mainline{8.Kf3} \dtmw{11},
            \restoregame{gbe4121} \mainline{8.Kxe3} \dtmw{},
            \restoregame{gbe4121} \mainline{8.c5} \dtmw{11}\color{black})

            \restoregame{gbe4121} \mainline{8.Bxe3 Ne5} 
            \storegame{gbe14121a}
            (\color{red}\restoregame{gbe14121a}\mainline{9.Rb3} \dtmw{19},
            \restoregame{gbe14121a} \mainline{9.Rxb4} \dtmw{13},
            \restoregame{gbe14121a} \mainline{9.Rc2} \dtmw{12},
            \restoregame{gbe14121a} \mainline{9.Rd2} \dtmw{23},
            \restoregame{gbe14121a} \mainline{9.Bd2} \dtmw{13},
            \restoregame{gbe14121a} \mainline{9.Bf2} \dtmw{20},
            \restoregame{gbe14121a} \mainline{9.Bd4} \dtmw{15},
            \restoregame{gbe14121a} \mainline{9.Bc5} \dtmw{12},
            \restoregame{gbe14121a} \mainline{9.Bxb6} \dtmw{19},
            \restoregame{gbe14121a} \mainline{9.Kd2} \dtmw{12},
            \restoregame{gbe14121a} \mainline{9.Kf2} \dtmw{12}\color{black})
            
            \subitem \varid{4}{1.2.1.1} \restoregame{gbe14121a} 
            \mainline{9.c5 Bxc5} \draw \justif 
            \mainline{10.Bxc5 Nc4 11.Rxb4 Rxb4 12. Bxb4}.

            \subitem \varid{4}{1.2.1.1} \restoregame{gbe14121a} 
            \mainline{9.fxe5 Bxe5} \draw \justif{10.c5 Rc6 
            11.Rxb4 Kd5}

      \item \varid{4}{1.2.2} \restoregame{gbe312a}
            \mainline{7.Kd3 Bc5} \storegame{gbe312aa}
            (\color{red}\restoregame{gbe312aa}\mainline{8.Rxb4} \dtmw{24},
            \restoregame{gbe312aa}\mainline{8.Nxb4} \dtmw{24},
            \restoregame{gbe312aa}\mainline{8.Nxd4+} \dtmw{24},
            \restoregame{gbe312aa}\mainline{8.Ne3} \dtmw{21},
            \restoregame{gbe312aa}\mainline{8.Bxb4} \dtmw{24},
            \restoregame{gbe312aa}\mainline{8.Bc3} \dtmw{11},
            \restoregame{gbe312aa}\mainline{8.Be3} \dtmw{21},
            \restoregame{gbe312aa}\mainline{8.Ke2} \dtmw{18}\color{black}) 
 
           \restoregame{gbe312aa}
            \mainline{8.Rb3 Kd6 9.Ke2 Ke6} \draw White is blocked and
            cannot do anything concrete in this position.
 
      \item \varid{4}{1.2.3} \restoregame{gbe312a} 
            \mainline{7.Rb3 Bc5} \draw this move simply transposes to
            variation \varid{4}{1.2.2}
    \end{itemize}

    \item \varid{4}{1.3} \restoregame{gbe31}\mainline{5.cxd4 cxd4}
          \draw the position is totally blocked Black just moves his
          King on e6 d5. 
          
        \item \varid{4}{1.4} \restoregame{gbe31}\mainline{5.Rb3 c4}  
          \storegame{gbe314}
          (\color{red}\restoregame{gbe314}\mainline{6.dxc4} \dtmw{24},
           \restoregame{gbe314}\mainline{6.cxd4} \dtmw{13},
           \restoregame{gbe314}\mainline{6.Nxd4} \dtmw{10}\color{black})

           \restoregame{gbe314}\mainline{6.Rb2 dxc3}
           \storegame{gbe314a}
           (\color{red}\restoregame{gbe314a}\mainline{7.dxc4} \dtmw{10}, 
           \restoregame{gbe314a}\mainline{7.Ke3} \dtmw{6}, 
           \restoregame{gbe314a}\mainline{7.Be3} \dtmw{6}\color{black}) 
        
           \restoregame{gbe314a}\mainline{7.Bxc3 cxd3+} 
           \storegame{gbe314b}
           (\color{red}\restoregame{gbe314b}\mainline{8.Kf3} \dtmw{15}, 
           \restoregame{gbe314b}\mainline{8.Ke3} \dtmw{19},
           \restoregame{gbe314b}\mainline{8.Kd2} \dtmw{13}\color{black})

           \restoregame{gbe314b} \mainline{8.Kxd3 Bxf4} 
           \storegame{gbe314c}
           (\color{red}\restoregame{gbe314c}\mainline{9.Rb3} \dtmw{33}, 
            \restoregame{gbe314c}\mainline{9.Bd4} \dtmw{29}, 
            \restoregame{gbe314c}\mainline{9.Ke2} \dtmw{26},
            \restoregame{gbe314c}\mainline{9.Be5} \dtmw{9},
            \restoregame{gbe314c}\mainline{9.Bf6} \dtmw{18},
            \restoregame{gbe314c}\mainline{9.Bd2} \dtmw{22}\color{black})

            \restoregame{gbe314c}
            \mainline{9.Nd4+ Nxd4}\storegame{gbe314d}
            (\color{red}\restoregame{gbe314d} \mainline{10.Rb3} \dtmw{9},
            \restoregame{gbe314d} \mainline{10.Rc2} \dtmw{10},
            \restoregame{gbe314d} \mainline{10.Rd2} \dtmw{11},
            \restoregame{gbe314d} \mainline{10.Re2+} \dtmw{9},
            \restoregame{gbe314d} \mainline{10.Rf2} \dtmw{17},
            \restoregame{gbe314d} \mainline{10.Bd2} \dtmw{14},
            \restoregame{gbe314d} \mainline{10.Kxd4} \dtmw{11}\color{black})

            \restoregame{gbe314d} \mainline{10.Bxd4 Rc6} \draw
            Black will place his King on d5 and White cannot progress.            

    \item \varid{4}{1.5} \restoregame{gbe31}\mainline{5.Kf2 dxc3}
          \storegame{gbe318}
         (\color{red}\restoregame{gbe318}\mainline{6.bxc5} \dtmw{4},
         \restoregame{gbe318}\mainline{6.d4} \dtmw{5},
         \restoregame{gbe318}\mainline{6.Rb3} \dtmw{4},
         \restoregame{gbe318}\mainline{6.Nd4} \dtmw{10},
         \restoregame{gbe318}\mainline{6.Ne3} \dtmw{10},
         \restoregame{gbe318}\mainline{6.Be3} \dtmw{7},
         \restoregame{gbe318}\mainline{6.Ke2} \dtmw{6},
         \restoregame{gbe318}\mainline{6.Ke3} \dtmw{6},
         \restoregame{gbe318}\mainline{6.Kf3} \dtmw{5}\color{black})

         \restoregame{gbe318}\mainline{6.Bxc3 cxb4} 
         \storegame{gbe318a}
         (\color{red}\restoregame{gbe318a}\mainline{7.d4} \dtmw{13}, 
         \restoregame{gbe318a}\mainline{7.Rb3} \dtmw{16}, 
         \restoregame{gbe318a}\mainline{7.Rxb4} \dtmw{23}, 
         \restoregame{gbe318a}\mainline{7.Nd4+} \dtmw{24}, 
         \restoregame{gbe318a}\mainline{7.Ne3} \dtmw{11}, 
         \restoregame{gbe318a}\mainline{7.Bf6} \dtmw{18}, 
         \restoregame{gbe318a}\mainline{7.Be5} \dtmw{13}, 
         \restoregame{gbe318a}\mainline{7.Ke2} \dtmw{12}, 
         \restoregame{gbe318a}\mainline{7.Ke3} \dtmw{13}, 
         \restoregame{gbe318a}\mainline{7.Kf3} \dtmw{13}\color{black})
         \begin{itemize}
           \item \varid{4}{1.5.1} \restoregame{gbe318a}
             \mainline{7.Bxb4 Nxb4} 
             \storegame{gbe318a1} 
             (\color{red}\restoregame{gbe318a1}\mainline{8.d4} \dtmw{11},
              \restoregame{gbe318a1}\mainline{8.Rb3} \dtmw{10},
              \restoregame{gbe318a1}\mainline{8.Rxb4} \dtmw{18},
              \restoregame{gbe318a1}\mainline{8.Nd4} \dtmw{13},
              \restoregame{gbe318a1}\mainline{8.Ne3} \dtmw{13},
              \restoregame{gbe318a1}\mainline{8.Ke2} \dtmw{12},
              \restoregame{gbe318a1}\mainline{8.Ke3} \dtmw{13},
              \restoregame{gbe318a1}\mainline{8.Kf3} \dtmw{12}\color{black})

              \restoregame{gbe318a1}\mainline{8.Nxb4 Bxb4}
              \storegame{gbe318a1a} 
              (\color{red}\restoregame{gbe318a1a}\mainline{9.d4} \dtmw{11},
              \restoregame{gbe318a1a}\mainline{9.Rb3} \dtmw{12},
              \restoregame{gbe318a1a}\mainline{9.Rc2} \dtmw{11},
              \restoregame{gbe318a1a}\mainline{9.Rd2} \dtmw{7},
              \restoregame{gbe318a1a}\mainline{9.Re2+} \dtmw{12},
              \restoregame{gbe318a1a}\mainline{9.Ke2} \dtmw{11},
              \restoregame{gbe318a1a}\mainline{9.Ke3} \dtmw{10},
              \restoregame{gbe318a1a}\mainline{9.Kf3} \dtmw{11}\color{black}) 
              \restoregame{gbe318a1a}\mainline{9.Rxb4 Kd5} \draw this
              Rook endgame is a draw. 

           \item \varid{4}{1.5.2} \restoregame{gbe318a}
              \mainline{7.Bd2 b3} \storegame{gbe318b}
              (\color{red}\restoregame{gbe318b}\mainline{8.Ne3} \dtmw{33},
              \restoregame{gbe318b} \mainline{8.Nb4} \dtmw{14},
              \restoregame{gbe318b} \mainline{8.Be3} \dtmw{23},
              \restoregame{gbe318b} \mainline{8.d4} \dtmw{14},
              \restoregame{gbe318b} \mainline{8.Bc3} \dtmw{14},
              \restoregame{gbe318b} \mainline{8.Ke3} \dtmw{21},
              \restoregame{gbe318b} \mainline{8.Bb4} \dtmw{11},
              \restoregame{gbe318b} \mainline{8.Nd4} \dtmw{12},
              \restoregame{gbe318b} \mainline{8.Kf3} \dtmw{12},
              \restoregame{gbe318b} \mainline{8.Ke2} \dtmw{12}\color{black})

              \restoregame{gbe318b} \mainline{8.Rxb3 Bc5+} 
              \storegame{gbe318c}
              (\color{red} \restoregame{gbe318c} \mainline{9.Ne3} \dtmw{23},
              \restoregame{gbe318c} \mainline{9.Nd4+} \dtmw{12}\color{black})

              \subitem * \varid{4}{1.5.2.1} \restoregame{gbe318c}
                         \mainline{9.Be3 Bxe3+} \draw the Black King 
                         will block the position on d5.

              \subitem * \varid{4}{1.5.2.2} \restoregame{gbe318c}
                         \mainline{9.d4 Nxd4} \draw 
                         \justif \mainline{10.Nxd4+ Nxd4 11.Rd3 b4
                           12.Be3 b3 13.Bxd4} the White Bishop must be
                         exchanged vs the b pawn and the Rook ending
                         is draw.  

              \subitem * \varid{4}{1.5.2.3} \restoregame{gbe318c}
                         \mainline{9.Ke2 Nd4+} \draw \justif
                         \mainline{10.Nxd4+ Bxd4} and Black locks the
                         position by \bmove{Kd5}. 

              \subitem * \varid{4}{1.5.2.4} \restoregame{gbe318c}
                         \mainline{9.Kf3 Kd5} \draw the reason has to
                         be seen in previous lines. Black exchange the
                         Bishop vs the Knight and the remaining
                         position is blocked. 

           \item \varid{4}{1.5.3} \restoregame{gbe318a}
             \mainline{7.Bd4 Nxd4} \storegame{gbe3183} 
             (\color{red}\restoregame{gbe3183}\mainline{8.Rb3} \dtmw{8},
             \restoregame{gbe3183}\mainline{8.Rxb4} \dtmw{10},
             \restoregame{gbe3183}\mainline{8.Nxb4} \dtmw{10},
             \restoregame{gbe3183}\mainline{8.Ne3} \dtmw{9},
             \restoregame{gbe3183}\mainline{8.Ke3} \dtmw{8}\color{black})

             \restoregame{gbe3183}\mainline{8.Nxd4+ Kd5}
             \storegame{gbe31383a}
             (\color{red}\restoregame{gbe31383a}\mainline{9.Rb3} \dtmw{9},
             \restoregame{gbe31383a}\mainline{9.Rxb4} \dtmw{10},
             \restoregame{gbe31383a}\mainline{9.Nb3} \dtmw{31},
             \restoregame{gbe31383a}\mainline{9.Nxb5} \dtmw{10},
             \restoregame{gbe31383a}\mainline{9.Nc6} \dtmw{11},
             \restoregame{gbe31383a}\mainline{9.Ne2} \dtmw{22},
             \restoregame{gbe31383a}\mainline{9.Ne6} \dtmw{12},
             \restoregame{gbe31383a}\mainline{9.Nf3} \dtmw{23},
             \restoregame{gbe31383a}\mainline{9.Rc2} \dtmw{11},
             \restoregame{gbe31383a}\mainline{9.Rd2} \dtmw{12},
             \restoregame{gbe31383a}\mainline{9.Re2} \dtmw{12},
             \restoregame{gbe31383a}\mainline{9.Ke2} \dtmw{11},
             \restoregame{gbe31383a}\mainline{9.Ke3} \dtmw{11},
             \restoregame{gbe31383a}\mainline{9.Kf3} \dtmw{11}\color{black})

               -  \varid{4}{1.5.3.1}\restoregame{gbe31383a}
                 \mainline{9.Nc2 Bc5+} \draw for the same reasons as
                 lines \varid{4}{1.5.2.1} / \varid{4}{1.5.2.2} /
                 \varid{4}{1.5.2.3} / \varid{4}{1.5.2.4}

               - \varid{4}{1.5.3.2}\restoregame{gbe31383a}
                 \mainline{9.Nxf5 Bc5+} \draw once again the Black
                 Bishop is exchanged vs the Knight and the remaining
                 Rook ending is draw. 
          \end{itemize}

           \item \varid{4}{1.5.4} \restoregame{gbe318a}
             \mainline{7.Nxb4 Bxb4} all moves but one lead
             to White checkmate \storegame{gbe318a4}(
             \color{red}
             \restoregame{gbe318a4}\mainline{8.d4} \dtmw{10},
             \restoregame{gbe318a4}\mainline{8.Rb3} \dtmw{17},
             \restoregame{gbe318a4}\mainline{8.Rxb4} \dtmw{11},
             \restoregame{gbe318a4}\mainline{8.Rc2} \dtmw{11},
             \restoregame{gbe318a4}\mainline{8.Rd2} \dtmw{10},
             \restoregame{gbe318a4}\mainline{8.Re2+} \dtmw{13},
             \restoregame{gbe318a4}\mainline{8.Bd2} \dtmw{11},
             \restoregame{gbe318a4}\mainline{8.Bd4} \dtmw{7},
             \restoregame{gbe318a4}\mainline{8.Be5} \dtmw{11},
             \restoregame{gbe318a4}\mainline{8.Bf6} \dtmw{12},
             \restoregame{gbe318a4}\mainline{8.Ke2} \dtmw{9},
             \restoregame{gbe318a4}\mainline{8.Ke3} \dtmw{10},
             \restoregame{gbe318a4}\mainline{8.Kf3} \dtmw{9}\color{black})
         \restoregame{gbe318a4}
         \mainline{8.Bxb4 Nxb4} \draw \storegame{gbe318a4a} 
         (\color{red}\restoregame{gbe318a4a}\mainline{9.d4} \dtmw{8},
         \restoregame{gbe318a4a}\mainline{9.Rb3} \dtmw{11},
         \restoregame{gbe318a4a}\mainline{9.Rc2} \dtmw{7},
         \restoregame{gbe318a4a}\mainline{9.Rd2} \dtmw{12},
         \restoregame{gbe318a4a}\mainline{9.Re2+} \dtmw{12},
         \restoregame{gbe318a4a}\mainline{9.Ke3} \dtmw{12},
         \restoregame{gbe318a4a}\mainline{9.Ke2} \dtmw{12},
         \restoregame{gbe318a4a}\mainline{9.Kf3} \dtmw{11}\color{black})
         
        \restoregame{gbe318a4a}
        \mainline{9.Rxb4}
         leads to the same drawn Rook endgame as \varid{4}{1.5.1}.
         \end{itemize}

    \item \varid{4}{1.6} \restoregame{gbe31}\mainline{5.Kf3 cxb4}
          \storegame{gbe416}
          (\color{red}\restoregame{gbe416}\mainline{6.Kf2} \dtmw{13},
          \restoregame{gbe416} \mainline{6.c4} \dtmw{27},
          \restoregame{gbe416} \mainline{6.Rxb4} \dtmw{17},
          \restoregame{gbe416} \mainline{6.Rb3} \dtmw{14},
          \restoregame{gbe416} \mainline{6.Be3} \dtmw{13},
          \restoregame{gbe416} \mainline{6.Ke2} \dtmw{14},
          \restoregame{gbe416} \mainline{6.Ne3} \dtmw{11}\color{black})
          \begin{itemize}
            \item \varid{4}{1.6.1} \restoregame{gbe416}
              \mainline{6.cxd4 b3} \storegame{gbe4161}
              (\color{red}\restoregame{gbe4161}\mainline{7.Kf2} \dtmw{15},
              \restoregame{gbe4161}\mainline{7.Ke2} \dtmw{11},
              \restoregame{gbe4161}\mainline{7.Ke3} \dtmw{14},
              \restoregame{gbe4161}\mainline{7.Bc3} \dtmw{19},
              \restoregame{gbe4161}\mainline{7.Bb4} \dtmw{11},
              \restoregame{gbe4161}\mainline{7.Be3} \dtmw{26},
              \restoregame{gbe4161}\mainline{7.Ne3} \dtmw{15},
              \restoregame{gbe4161}\mainline{7.Nb4} \dtmw{14}\color{black})
              
              \subitem
              \varid{4}{1.6.1.1}\restoregame{gbe4161}\mainline{7.Rxb3
                b4} \storegame{gbe4161a}
              (\color{red}\restoregame{gbe4161a}\mainline{8.Bxb4}\dtmw{35},
              \restoregame{gbe4161a} \mainline{8.Rc3} \dtmw{17},
              \restoregame{gbe4161a} \mainline{8.Rxb4} \dtmw{20},
              \restoregame{gbe4161a} \mainline{8.Nxb4} \dtmw{15},
              \restoregame{gbe4161a} \mainline{8.Ne3}\dtmw{11}\color{black})
              \draw White cannot progress wihtout giving a piece or
              moving \wmove{d5+} after which the Black King blocks the
              position. If White moves around Black simply plays his
              Rook on b5-b6. \justif \restoregame{gbe4161a}
              \mainline{8.d5+ Kxd5 9.Ne3+ Ke6}
              \storegame{gbe4161b} 
              (\color{red}\restoregame{gbe4161b}\mainline{10.Ke2}\dtmw{11},
              \restoregame{gbe4161b}\mainline{10.Nd5} \dtmw{11},
              \restoregame{gbe4161b}\mainline{10.Rc3} \dtmw{14},
              \restoregame{gbe4161b}\mainline{10.Bc3} \dtmw{33},
              \restoregame{gbe4161b}\mainline{10.Rb2} \dtmw{33},
              \restoregame{gbe4161b}\mainline{10.Rxb4} \dtmw{20},
              \restoregame{gbe4161b}\mainline{10.Bxb4} \dtmw{17},
              \restoregame{gbe4161b}\mainline{10.Nxf5} \dtmw{24},
              \restoregame{gbe4161b}\mainline{10.Nc4} \dtmw{14},
              \restoregame{gbe4161b}\mainline{10.Kf2} \dtmw{29},
              \restoregame{gbe4161b}\mainline{10.d4} \dtmw{11}
              \color{black})  
              and the only non losing line is to repeat with 
              \restoregame{gbe4161b}\mainline{10.Nc2}. 
 
              \subitem
              \varid{4}{1.6.1.2}\restoregame{gbe4161}\mainline{7.d5+}
              \draw see line \varid{4}{1.6.1.1}.

            \item \varid{4}{1.6.2} \restoregame{gbe416}
              \mainline{6.cxb4 Kd5} \draw position is totally
              blocked. 
            \item \varid{4}{1.6.3} \restoregame{gbe416}
              \mainline{6.Nxd4+ Nxd4} \draw \justif \mainline{7.cxd4}
              Black puts his King on d5 and the White position is
              totally blocked. 
              
            \item \varid{4}{1.6.4} \restoregame{gbe416}
              \mainline{6.Nxb4 Nxb4} \draw \justif \mainline{7.cxb4
                Kd5} is similar to line \varid{4}{1.6.2}. 
          \end{itemize}

  \end{itemize}

  \item \varid{4}{2} \restoregame{gbe3}\mainline{4.cxd4 cxd4}
        \draw see line \varid{4}{3}. 

  \item \varid{4}{3} \restoregame{gbe3}\mainline{4.Kf2 b4} \draw
        the position is totally blocked and Black can just move his
        King on e6-f6 \justif \mainline{5.cxb4 cxb4 6.Kf3 Ke6} etc. 

  \item \varid{4}{4} \restoregame{gbe3}\mainline{4.Kf3 b4} \draw for
        the same reasons as in line \varid{4}{3}.
   
\end{itemize}
\restoregame{gbe3}
\mainline{4.c4 b4} \draw see line \varid{1}{5.2}
   \subsection{White moves \wmove{Nb4}}
\gardnerstart
\mainline{1.Nb4 cxb4} \dtmw{21} White is a piece down. 
   
   \subsection{White moves \wmove{Nd4}}
\gardnerstart
\mainline{1.Nd4 exd4} \dtmw{25} White is a piece down.

\section{Conclusion}

The game-theoretical value of Gardner's chess has been proved to be a
draw. The proof was done in a semi-automated way in which humans were
guiding the engine.  The authors were 'pushing' lines for which it was
thought that the exact distance to checkmate could be computed and
backtracked once leaves were showing perfect distance to
checkmate. This meta-algorithm leads to a very asymmetric way of
selecting moves. For instance, when a position is thought to be
decidable as a White win, very few time is spent on White decision
nodes (since we 'know' the game to be won more or less no matter
what). The idea is that enormous time and energy can be saved when
the game theoretic value of a position, rather than the most precise
move or the shortest path to checkmate, is looked for. Indeed, when a
game is thought to be winning, e.g. for White, one has only to provide
one forced line (even if it is not the 'best' one) and thus can avoid
exhaustive search at White decision nodes. It can be seen as a form of
meta-negascout \cite{Fish81}. Nevertheless it is very different in
the sense that the process is very asymmetric and guided by the fact
that the overall evaluation of the position is known.

This procedure can be fully automated and tuned to some given degree
of precision (basically what is the threshold after which a position
is considered as decided). For future works we plan to implement it
and test it for larger chess variants in order to compute their game
theoretic values. Other games could also be considered.

\subsection*{Acknowledgments}
  We thank Francois Challier and Philippe Virouleau for their
technical and technological help.  

\bibliographystyle{alpha}

\begin{thebibliography}{RCK{\etalchar{+}}10}

\bibitem[All94]{Alis94}
V.~Allis.
\newblock {\em Searching for Solutions in Games and Artificial Intelligence}.
\newblock PhD thesis, University of Limburg, Maastricht, The Netherlands, 1994.

\bibitem[Fis81]{Fish81}
J.~P. Fishburn.
\newblock {\em Analysis of Speedup in Distributed Algorithms}.
\newblock PhD thesis, University of Winconsin, Madison, 1981.

\bibitem[Kry04]{Kryu33}
K.~Kryukov.
\newblock 3 $\times$ 3 chess.
\newblock website : http://kirr.homeunix.org/chess/3x3-chess/, 2004.

\bibitem[Kry09]{Kryu34}
K.~Kryukov.
\newblock 3 $\times$ 4 chess.
\newblock website : http://kirr.homeunix.org/chess/3x4-chess/, 2009.

\bibitem[Kry11]{Kryu44}
K.~Kryukov.
\newblock 4 $\times$ 4 chess.
\newblock website : http://kirr.homeunix.org/chess/4x4-chess/, 2011.

\bibitem[Ltd13]{Lomonosov}
Convetka Ltd.
\newblock Lomonosov endgame tablebases.
\newblock website : http://chessok.com/?page\_id=27966, 2013.

\bibitem[Pri07]{Prit07}
D.~B. Pritchard.
\newblock {\em The Classified Encyclopedia of Chess Variants}.
\newblock J. Beasley, 2007.

\bibitem[Pro12]{Pro12}
F.~Prost.
\newblock On the impact of information technologies on society: an historical
  perspective through the game of chess.
\newblock In A.~Voronkov, editor, {\em Turing-100}, volume~10 of {\em EPIC},
  pages 268--277. EasyChair, 2012.

\bibitem[RCK{\etalchar{+}}10]{Stockfish}
T.~Romstad, M.~Costalba, J.~Kiiski, D.~Yang, S.~Spitaleri, and J.~Ablett.
\newblock Stockfish.
\newblock web site: http://stockfishchess.org/, 2010.

\bibitem[SBB{\etalchar{+}}07]{SchaBurBjorKis07}
J.~Schaeffer, N.~Burch, Y.~Bjornsson, A.~Kishimoto, M.~Muller, R.~Lake, P.~Lu,
  and S.~Sutphen.
\newblock Checkers is solved.
\newblock {\em Nature}, 317(5844):1518--1522, 2007.

\end{thebibliography}
\newcommand{\etalchar}[1]{$^{#1}$}

\end{document}